\newif\ifFull\Fulltrue

\ifFull
\newcommand{\full}[2]{#1}
\else
\newcommand{\full}[2]{#2}
\fi

\documentclass[nonacm,table,hyphens,pagebackref,acmsmall,screen]{acmart}
\setcopyright{rightsretained} 
\startPage{1}

\usepackage{xspace} 
\usepackage{amsmath,amsfonts}

\usepackage{tabu}
\usepackage{dcolumn} 
\usepackage{booktabs}
\usepackage{enumitem}
\usepackage{multirow}
\usepackage{wrapfig}

\usepackage{graphicx}
\usepackage{tikz}
\usepackage{pgfplots}

\usepackage{afterpage}
\usepackage{pdflscape}
\usepackage{caption}

\usepackage[frozencache,cachedir=minted]{minted}
\definecolor{mintedConstant}{RGB}{136,0,0} 
\usepackage{etoolbox}
\makeatletter
\AtBeginEnvironment{minted}{\dontdofcolorbox}
\def\dontdofcolorbox{\renewcommand\fcolorbox[4][]{##4}}
\makeatother

\usepackage{listings}
\usepackage{lstlangcoq}
\usepackage{subcaption}
\usepackage[ruled]{algorithm2e}

\usepackage{hyperref}
\usepackage[nameinlink,capitalise,noabbrev]{cleveref}
\usepackage{scalerel}
\usepackage{rotating}
\usepackage{soul}
\usepackage{float}

\usetikzlibrary{arrows,bending,calc,positioning,shadows,shapes}

\usepackage{array}
\usepackage{stmaryrd}

\newfloat{dummy}{htb}{dmy}

\newcommand{\parhead}[1]{\vspace{5pt plus 2pt minus 4pt}\par\noindent\textbf{#1}\hspace{.4em plus .2em minus .2em}} 

\definecolor{darkviolet}{HTML}{9400D3}
\definecolor{darkgreen}{rgb}{0,0.65,0}

\newcommand{\instr}[1]{\texttt{#1}}
\newlength{\mintednumbersep}
\AtBeginDocument{%
  \sbox0{\tiny00}%
  \setlength\mintednumbersep{\parindent}%
  \addtolength\mintednumbersep{-\wd0}%
}

\newcommand{\cryptopt}{\textsf{CryptOpt}\xspace}
\newcommand{\ots}{off-the-shelf\xspace}
\newcommand{\gm}{G.M.\xspace} 

\newcommand{\fiat}{\textsf{Fiat Cryptography}\xspace}
\newcommand{\fiatir}{\textsf{Fiat~IR}\xspace}
\newcommand{\assemblyline}{\textsf{AssemblyLine}\xspace}
\newcommand{\assembly}{assembly\xspace}
\newcommand{\xassembly}{x86-64~\assembly}
\newcommand{\nistp}{NIST P-256\xspace}

\newcommand{\cnothing}{c_{\scaleto{\varnothing}{4pt}}}

\newcommand{\rls}{{RLS}\xspace}
\newcommand{\barheur}{Bet-and-Run\xspace}
\newcommand{\approachRRR}{{R3-validation}\xspace}

\newcommand{\budgett}{200\,000\xspace}    
\newcommand{\budgetk}{20\xspace}          
\newcommand{\budgetpercent}{10\%\xspace}  
\newcommand{\budgettone}{20\,000\xspace}  
\newcommand{\budgetttwo}{180\,000\xspace} 
\newcommand{\budgeteach}{1\,000\xspace}      
\newcommand{\parallelruns}{three\xspace}

\newcommand{\cf}{\texttt{CF}}
\newcommand{\of}{\texttt{OF}}
\newcommand{\fadd}{\textsc{Add}\xspace}
\newcommand{\fmul}{\textsc{Mul}\xspace}
\newcommand{\gccVersion}{12.1.0\xspace}
\newcommand{\clangVersion}{15.0.6\xspace}

\SetCommentSty{mycommfont}
\newcommand{\colA}{blue} 
\newcommand{\colB}{red} 
\newcommand{\colC}{darkgreen} 
\newcommand{\colD}{orange} 
\newcommand{\colE}{violet} 
\newcommand{\colF}{cyan} 
\newcommand{\colG}{olive} 

\newcommand{\docolor}[2]{{\setlength{\fboxsep}{0pt}%
  \ifthenelse{\equal{#1}{A}}{\colorbox{green!15}{#2}}{%
  \ifthenelse{\equal{#1}{B}}{\colorbox{blue!15}{#2}}{%
  \ifthenelse{\equal{#1}{C}}{\colorbox{yellow!15}{#2}}{%
  \ifthenelse{\equal{#1}{D}}{\colorbox{red!15}{#2}}{%
    #1#2%
  }}}}%
}}%
{
\catcode`\!\active

}

\usepackage[utf8]{inputenc}
\usepackage{newunicodechar}
\newunicodechar{λ}{$\lambda$}
\newunicodechar{∀}{$\forall$}
\newunicodechar{⌊}{$\lfloor$}
\newunicodechar{⌋}{$\rfloor$}

\newcolumntype{d}[1]{D{.}{.}{#1}}
\newcommand\mc[1]{\multicolumn{1}{c}{#1}} 

\begin{document}

\title[\cryptopt: Verified Compilation with Randomized Program Search for Cryptographic Primitives]{\cryptopt: Verified Compilation with Randomized Program Search for Cryptographic
Primitives {\large (full version)}}

\newcommand{\uoa}{%
\affiliation{
  \institution{University of Adelaide}
  \country{Australia}
}}
\newcommand{\MIT}{%
\affiliation{
  \institution{Massachusetts Institute of Technology}
  \country{USA}
}}
\newcommand{\Stanford}{%
\affiliation{
  \institution{Stanford University}
  \country{USA}
}}
\newcommand{\gatech}{%
\affiliation{
  \institution{Georgia Institute of Technology}
  \country{USA}
}}
\newcommand{\monash}{%
\affiliation{
  \institution{Monash University}
  \country{Australia}
}}
\newcommand{\melbuni}{%
\affiliation{
  \institution{The University of Melbourne}
  \country{Australia}
}}
\newcommand{\rub}{%
\affiliation{
  \institution{Ruhr University Bochum}
  \country{Germany}
}}

\author{Joel Kuepper}\uoa
\author{Andres Erbsen}\MIT
\author{Jason Gross}\MIT
\author{Owen Conoly}\MIT
\author{Chuyue Sun}\Stanford
\author{Samuel Tian}\MIT
\author{David Wu}\uoa
\author{Adam Chlipala}\MIT
\author{Chitchanok Chuengsatiansup}\melbuni
\author{Daniel Genkin}\gatech
\author{Markus Wagner}\monash
\author{Yuval Yarom}\rub

\renewcommand{\shortauthors}{Kuepper,
Erbsen, 
Gross, 
Conoly, 
Sun, 
Tian, 
Wu,
Chlipala, 
Chuengsatiansup, 
Genkin, 
Wagner, 
Yarom}

\begin{abstract}
  \vfill
Most software domains rely on compilers to translate high-level code to multiple different machine languages, with performance not too much worse than what developers would have the patience to write directly in assembly language.
However, cryptography has been an exception, where many performance-critical routines have been written directly in assembly (sometimes through metaprogramming layers).
Some past work has shown how to do formal verification of that assembly, and other work has shown how to generate C code automatically along with formal proof, but with consequent performance penalties vs. the best-known assembly.
We present \cryptopt, the first compilation pipeline that specializes high-level cryptographic functional programs into assembly code significantly faster than what GCC or Clang produce, with mechanized proof (in Coq) whose final theorem statement mentions little beyond the input functional program and the operational semantics of \xassembly.
On the optimization side, we apply randomized search through the space of assembly programs, with repeated automatic benchmarking on target CPUs.
On the formal-verification side, we connect to the \fiat framework (which translates functional programs into C-like IR code) and extend it with a new formally verified program-equivalence checker, incorporating a modest subset of known features of SMT solvers and symbolic-execution engines.
The overall prototype is quite practical, e.g.\ producing new fastest-known implementations of finite-field arithmetic for both Curve25519 (part of the TLS standard) and the Bitcoin elliptic curve secp256k1 for the Intel $12^{th}$ and $13^{th}$ generations.

\vfill\vfill\vfill
\end{abstract}

\maketitle

\clearpage
$\ $
\vspace{4em}
\section{Introduction}\label{s:introduction}

Being a foundational pillar of computer security, cryptographic software needs to achieve three often-competing aims.
First, being security-critical, the software needs to be correct and protected from implementation attacks.
Second, because it is used frequently, it needs to be efficient.
Third, for migration to new architectures, the software needs to be portable.
Implementations of cryptographic code, therefore, must strike a trade-off between these needs.
Implementations that aim for portability tend to use high-level languages, such as Java or C. These allow for easy maintenance and are essentially platform-independent, assuming the existence of suitable development tools like compilers and assemblers.

At the same time, compiler-based code generation can be a double-edged sword.
First, compiler-produced cryptographic code tends to underperform when compared to hand-optimized
code~\cite{panda,curve41417,kummer,ntruprime,mlwrprf,mcbits,qcbits,Kannwischer2019,sandy2x}, typically written directly in the platform's \assembly language. 
Beyond slower performance, compilers are typically not designed for maintaining security properties. In particular, compilation bugs could result in incorrect code~\cite[\S V-B]{ErbsenPGSC19},
while overly aggressive optimizations may even strip side-channel protections~\cite{DSilvaPS15,Gilles18,Kaufmann16}.

We note that the difficulties compilers have when operating over cryptographic code are not caused by high code complexity. In fact, cryptographic code tends to be simpler than a typical program code,
due to its avoidance of data-dependent control flows and memory-access patterns for reasons of side-channel resistance. Ironically, compiler optimization passes often 
focus on control flow, as it offers higher impact than fine-tuning straight-line code~\cite{AhoSU86}. 

Instead, the cause is that such code tends to be simpler than a typical program code, and this simplicity deprives the compiler of optimization opportunities.
At the same time, we observe that such simplicity may offer opportunities for using strategies not commonly exploited by compilers, such as reordering arithmetic operations within
a basic block or exchanging machine instructions with semantically equivalent machine instructions.
Thus, our work starts from the question:

\medskip
\begin{centering}
\emph{How can we exploit the simplicity of cryptographic primitives in order to generate efficient and provably correct implementations of cryptographic functions?}
\end{centering}

\subsection{Our Contribution}
\begin{wrapfigure}{r}{0.3\textwidth}
    \centering
    \tikzset{every picture/.style={line width=0.75pt}} 
\begin{tikzpicture}[-latex, yscale=.7,xscale=.7]
    \scriptsize
    \tikzstyle{codeblock} = [draw=black, inner sep=.4em, text width=5.5em, minimum height=3em, align=center];
    \tikzstyle{lang} = [inner sep=.4em, text width=5.5em, minimum height=2em, align=center, fill=blue!20];

    \node[codeblock, text width=7em] (fiat) at (0,0) {Fiat\\Cryptography};
    \node[text width=5em, align=center] (input) [left=1.5em of fiat] {Field\\Parameters};
    \draw[-latex] (input.east) -- (fiat.west);
    \node[lang, below=3em of input] (ir) {\fiatir};
    \draw[-latex] (fiat.south) -- ++(0,-.4) -| (ir.north);
    \node[codeblock,right=1.5em of ir] (opt)  {Optimizer};
    \node[codeblock, below=10em of input] (checker)   {Checker};

    \draw[-latex] (ir.east) -- (opt.west);
    \draw[-latex] ([xshift=1em]ir.south west) -- ([xshift=1em]checker.north west);

    \node[lang, below=1.8em of opt] (assembly) {Assembly};

    \draw[-latex] (opt.south) --  (assembly.north);
    \draw[-latex] (assembly.west) -|  ([xshift=-1em]checker.north east);

\end{tikzpicture}
    \caption{Integration design. Boxes represent individual components, arrows represent data flow, and filled boxes represent files. \emph{Optimizer} and \emph{Checker} are results
    of this work. \fiat and \emph{Checker} are formally verified.\label{fig:contrib}}
\end{wrapfigure}
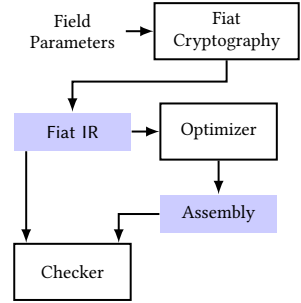

We present \cryptopt, a new code generator that produces highly efficient code tailored to the architecture it runs on.
The task is split between \emph{finding} performant program variants and \emph{checking} that they have preserved program behavior.
The former works via randomized search, and the latter works via a formally verified program-equivalence checker that should be applicable even with other optimization strategies.
As a result, the randomized-search procedures need not be trusted, and, when we compose them with the \fiat Coq-verified compiler~\cite{ErbsenPGSC19} and our new equivalence checker (as shown in \cref{fig:contrib}),
we get end-to-end functional-correctness proofs for fast assembly code --
faster than any code demonstrated for the cryptographic algorithms we study, when compiling automatically from high-level programs as we do.

To \emph{find} performant machine-code variants, instead of relying on human heuristics, \cryptopt represents code generation as a combinatorial optimization problem.
That is, \cryptopt considers a solution space that consists of machine-code implementations of the target function and uses techniques from the randomized-search domain
to seek the best-performing implementation.

To optimize, \cryptopt first chooses an arbitrary correct implementation of the target function.
It then mutates the implementation by either changing the instruction(s) that implements a certain operation or changing the order of operations.
If the mutated implementation is not worse than the original, the mutated implementation is kept; otherwise it is discarded.

Instead of trying to predict code performance like in work by \citet{Schkufza0A13,JoshiNR02}, \cryptopt measures actual execution time. 
This approach is important because it avoids inaccuracies inherent in hardware models and
allows \cryptopt to tailor produced code to the target processor while treating the processor itself as an opaque unit.
A particular advantage of the approach is that once manufacturers release new hardware, which changes e.g. pipelining effects or caching behavior, \cryptopt adapts to it
automatically without requiring manual adaptation of hardware models.
This \emph{find-and-optimize} approach is implemented by the \emph{Optimizer} component in \cref{fig:contrib}.

To \emph{check} that generated code is correct, \cryptopt integrates with \fiat~\cite{ErbsenPGSC19}.
That framework, implemented in Coq, already generates low-level IR programs proven to preserve behavior of high-level functional programs, and it has been adopted by all major web browsers and mobile platforms for small but important parts of their TLS implementations, so there is high potential for impact improving performance further without sacrificing formal rigor.
\cryptopt begins with \fiatir programs and generates \xassembly code.
To avoid needing to model the randomized-search process, we instead build and prove an equivalence checker, which can compare programs across \fiatir and \xassembly.
Its two main pieces implement modest subsets of features known from SMT solvers and symbolic-execution engines.
From SMT solvers, we take an E-graph data structure~\cite{DetlefsNS05} to canonicalize logical expressions via rewrite rules.
From symbolic-execution engines, we take maintenance of symbolic states that tie registers and symbolic memory locations to logical expressions known to the solver.
Thanks to their combined proof in Coq, none of these details need to be trusted.
This \emph{checking} is done by the \emph{Checker} component in \cref{fig:contrib}.

We evaluate \cryptopt using finite fields with nine different prime moduli.
We use \fiat to generate the \fiatir for the multiply and square operations for these fields.
We then use the \cryptopt optimizer to generate \xassembly code for these operations and the equivalence checker to verify that the code matches the \fiatir.
The produced \xassembly code achieves a mean speedup of 1.74  compared to GCC~\gccVersion (speedup 1.40 against Clang~\clangVersion),
across ten different x86-64 platforms (four AMD, six Intel). 

We further evaluate the \cryptopt optimizer as a stand-alone code generator.  
For that, we create an input function from the C code of \texttt{libsecp256k1}~\cite{libsecp256k1}, feed it into the \cryptopt optimizer, and obtain an average speedup of 1.04  against the hand-optimized assembly code in \texttt{libsecp256k1}.
As we do not have \fiatir code matching \texttt{libsecp256k1}, we cannot use the equivalence checker to verify the code we produce.

\clearpage
\subsection{Summary of Contributions}
In summary, we make the following contributions in this paper:
\begin{itemize}[nosep, leftmargin=*]
  \item We present \cryptopt, a code generator that relies on combinatorial optimization instead of compiler heuristics for producing highly efficient code (\cref{s:randomsearch}).  This broad approach is shown to apply to larger routines than past work has tackled successfully, while also being integrated with formal verification for the first time.
  \item We demonstrate that a relatively modest \fiat extension (\cref{s:equivalence}) suffices to enable integration with a wide range of backend compiler heuristics. We implemented and verified Coq functional programs with a minimal subset of well-known features from SMT solvers and symbolic-execution engines, leading to a single extractable compiler that checks assembly files for semantic equivalence with high-level functional programs. To our knowledge, this is the first such equivalence checker with mechanized proof from first principles.
  \item We generate formally verified high-performance cryptographic code optimized for ten CPU architectures, obtaining considerable speedups over GCC and Clang (\cref{s:evaluation}).
\end{itemize}

\medskip
The source code for \cryptopt is available at \url{https://github.com/0xADE1A1DE/CryptOpt}.

\section{Background}
In this section we present basic background required for the rest of the paper.

\subsection{Random Local Search}\label{s:rls}
A combinatorial optimization problem aims to find an optimal solution (e.g.\ one that minimizes a given \emph{objective function}) within a discrete set of candidate solutions.
Random Local Search (\rls)~\cite{AugerD11,doerr2019theory} is a simple optimization strategy that is often efficient and effective in finding local optima.
A run of \rls starts from a random candidate solution.
It then applies a random \emph{mutation} to the current candidate solution, generating another solution within the possible solutions set.
If the mutation improves (or does not deteriorate) the solution, the mutated solution is kept, replacing the current candidate solution.
Otherwise, the mutation is discarded, and the current candidate solution remains unchanged.
This mutation and evaluation repeats until some predefined termination condition is satisfied.

\rls is often highly sensitive to the initial conditions, i.e.\ the  candidate solution it starts from. 
To address such erratic behavior, the simple \barheur heuristic~\cite{FischettiM14betandrun, Weise19generalBetAndRun} turns the sensitivity to the initial conditions into an advantage by employing multiple runs.  
A typical use of \barheur
starts with multiple independent runs of RLS, each optimizing for a predefined number of mutations.
After this initialization step, the algorithm selects the best run and lets this run continue optimizing from that step, stopping when a total number of mutations is explored.

\subsection{Finite-Field Arithmetic for Cryptography}

We focus on elliptic-curve cryptography (ECC), which is used widely in Internet standards like TLS.
It involves certain geometric aspects that are orthogonal to our tooling, which supports arithmetic modulo large prime numbers, otherwise known as finite-field arithmetic (FFA).

Because the FFA is a performance-critical component in the implementations, many tend to implement it by hand.
For instance, targeting different architectures, the ubiquitous cryptographic library OpenSSL~\cite{openssl} has many hand-optimized implementations for Curve25519 and \nistp, which are both
well-known instantiations of ECC.
The Bitcoin blockchain uses yet another curve called secp256k1 for their block signatures.
Its core library \texttt{libsecp256k1} contains hand-optimized code for the field arithmetic as well as a C implementation used as a fallback in architectures for which no optimized version exists.

FFA is not trivial to implement.
In particular, a field element is typically represented by multiple limbs using several CPU registers, and thus every field operation requires multiple CPU cycles.
However, these implementations tend to be straight-line code, heavy on arithmetic rather than control flow, leading to ineffective optimization by standard compilers.
Human experts instead manually apply simultaneous instruction selection, instruction scheduling, and register allocation, which, going well beyond capabilities of \ots C compilers, should take into account microarchitecture details such as macro-op fusion~\cite{CelioDPA16, Ronen04}, cache prediction~\cite{Samantika06, Lepak00,
Hooker13}, cache-replacement policies~\cite{Vila20}, and other (potentially undocumented) microarchitectural choices.

\subsection{\fiat}

\citet{ErbsenPGSC19} present the \fiat framework, which translates descriptions of field arithmetic
into code with Coq proof of functional correctness.
The starting point is a library of functional programs that are used as templates for generating code for performing operations in finite fields.
These functional programs, which have been proven correct, can be specialized with a specific prime order for generating an intermediate representation (\fiatir) of the code that performs field arithmetic operations for the required field.

In execution, \fiat selects the functional program to specialize for the required field size and produces provably correct \fiatir.
It then uses one of the available backends to process the \fiatir code and produce an implementation in one of the supported languages, including C, Java, and Zig.

\subsection{Equivalence Checking}

\hyphenation{Comp-Cert}
Formally proved compilers are the gold standard to address concerns of optimization soundness.
For instance, CompCert~\cite{CompCertBackend_JAR09,LeroyBKSPF16} was proved as a correct C compiler using the Coq theorem prover, which we also rely on.
However, proving a whole compiler can be very labor-intensive, and thus it is often appealing to prove a \emph{checker} for compiler outputs, known as a translation validator.
For instance, CompCert was extended in that way~\cite{Tristan}.
To date, however, the formally proved translation validators have not incorporated reasoning with algebraic properties of arithmetic, as we found we needed in \cryptopt.

In contrast, translation validation with SMT solvers uses rich reasoning to prove equivalence between the high-level input program and the obtained low-level output. 
The Alive project~\cite{Alive2} for LLVM is a good example.
Compared to work with formally proved translation validators, Alive and similar tools include much larger trusted bases, for instance including a full SMT solver like Z3~\cite{z3}.
Some SMT solvers have been extended to produce proofs that can be checked in tools like Coq, as in SMTCoq~\cite{smtcoq}.

In our experience, these tools hit performance bottlenecks when working with large bitvectors.
Even if we imagine those issues as fixed some day, there are still benefits to creating a customized checker, keeping just the relevant aspects of SMT.
A benefit of a slimmed-down custom prover is that it becomes more feasible to prove the prover itself, rather than just a checker for its outputs, which improves
performance and reduces surprise from proof-generation bugs.

\subsection{The E-graph}

Following SMT-solver conventions~\cite{DetlefsNS05}, our E-graphs include nodes for equivalence classes of symbolic expressions, in addition to the edges representing subterm relationships.
Each node is configured to present the most compact representation of its equivalence class.
For instance, whenever a node becomes provably equal to a constant, it is labeled with that constant, without outgoing edges.
When a node is most succinctly expressed as a sum, it is labeled with a ``+'' operator and has edges to the other nodes being added.

\begin{wrapfigure}{r}{0.66\linewidth}
\newcommand{\esp}{\!\!\;\!}

\begin{tikzpicture}[scale=0.66,font=\relscale{.66}]

\draw(0,0.5) -- (14,0.5);
\draw(7.2,-3.5) -- (7.2,3.7);


\draw(0.5,3.25) circle (2mm) node{1};
\draw(1.25,3.25) node[anchor=west]{\textbf{\underline{Add:}} $\; x+z+(y \gg 9)+y$};

\draw[-stealth] (1.25,1.75) -- (1.1,1.3);
\draw[-stealth] (2.75,1.75) -- (2.9,1.3);
\draw[fill=white] (2,2) ellipse (10mm and 5mm);
\draw(2,2.15) node{$2\esp: +(0,1)$};
\draw(2,1.8) node{$x+y$};
\draw(1,1) ellipse (4mm and 3mm) node{$0\esp:x$};
\draw(3,1) ellipse (4mm and 3mm) node{$1\esp:y$};

\draw(4,2) ellipse (4mm and 3mm) node{$3\esp:z$};

\draw[-stealth] (5.75,1.75) -- (3.4,1.1);
\draw[-stealth] (6.25,1.75) -- (6.5,1.3);
\draw[fill=white] (6,2) ellipse (10mm and 5mm);
\draw(6,2.15) node{$5\esp: \gg (1,4)$};
\draw(6,1.8) node{$y \gg 9$};
\draw(6.5,1) ellipse (4mm and 3mm) node{$4\esp:9$};


\begin{scope}[shift={(7,0)}]

\draw(0.5,3.25) circle (2mm) node{2};
\draw(1.25,3.25) node[anchor=west]{\textbf{\underline{Add:}} $\; x+z+(y \gg 9)+y$};

\draw[-stealth,color=blue] (2.3,3.1) to[out=200,in=120] (0.8,1.3);
\draw[-stealth,color=blue] (3.6,3.05) -- (3.15,1.3);
\draw[-stealth,color=blue] (2.95,3.1) -- (3.7,2.2);
\draw[-stealth,color=blue] (4.3,3.1) to[out=270,in=170] (6.1,1);
\draw[-stealth,color=blue] (5.0,3.05) to[out=260,in=10] (3.3,1.2);

\draw[-stealth] (1.25,1.75) -- (1.1,1.3);
\draw[-stealth] (2.75,1.75) -- (2.9,1.3);
\draw[fill=white] (2,2) ellipse (10mm and 5mm);
\draw(2,2.15) node{$2\esp: +(0,1)$};
\draw(2,1.8) node{$x+y$};
\draw(1,1) ellipse (4mm and 3mm) node{$0\esp:x$};
\draw(3,1) ellipse (4mm and 3mm) node{$1\esp:y$};

\draw(4,2) ellipse (4mm and 3mm) node{$3\esp:z$};

\draw[-stealth] (5.75,1.75) -- (3.4,1.0); %
\draw[-stealth] (6.25,1.75) -- (6.5,1.3);
\draw[fill=white] (6,2) ellipse (10mm and 5mm);
\draw(6,2.15) node{$5\esp: \gg (1,4)$};
\draw(6,1.8) node{$y \gg 9$};
\draw(6.5,1) ellipse (4mm and 3mm) node{$4\esp:9$};

\end{scope}


\begin{scope}[shift={(0,-3.5)}]

\draw(0.5,3.25) circle (2mm) node{3};
\draw(1.25,3.25) node[anchor=west]{\textbf{\underline{Add:}} $\; x+z+(y \gg 9)+y$};

\draw[-stealth,color=blue] (2.3,3.1) to[out=200,in=120] (0.8,1.3);
\draw[-stealth,color=blue] (3.6,3.05) -- (3.15,1.3);
\draw[-stealth,color=blue] (2.95,3.1) -- (3.7,2.2);
\draw[-stealth,color=blue] (4.3,3.1) to[out=270,in=170] (6.1,1);
\draw[-stealth,color=blue] (5.0,3.05) to[out=260,in=10] (3.3,1.2);

\draw[-stealth] (1.25,1.75) -- (1.1,1.3);
\draw[-stealth] (2.75,1.75) -- (2.9,1.3);
\draw[fill=white] (2,2) ellipse (10mm and 5mm);
\draw(2,2.15) node{$2\esp: +(0,1)$};
\draw(2,1.8) node{$x+y$};
\draw(1,1) ellipse (4mm and 3mm) node{$0\esp:x$};
\draw(3,1) ellipse (4mm and 3mm) node{$1\esp:y$};

\draw(4,2) ellipse (4mm and 3mm) node{$3\esp:z$};

\draw[-stealth] (5.75,1.75) -- (3.4,1.0); %
\draw[-stealth] (6.25,1.75) -- (6.5,1.3);
\draw[fill=white] (6,2) ellipse (10mm and 5mm);
\draw(6,2.15) node{$5\esp: \gg (1,4)$};
\draw(6,1.8) node{$y \gg 9$};
\draw(6.5,1) ellipse (4mm and 3mm) node{$4\esp:9$};

\draw[-stealth,color=red,thick] (3.9,3.4) to[out=60,in=100] (6,2.5);

\end{scope}


\begin{scope}[shift={(7,-3.5)}]

\draw(0.5,3.25) circle (2mm) node{4};
\draw(3.5,3.25) ellipse (17mm and 6mm);
\draw(3.5,3.5) node{$6\esp: +(0,1,3,5)$};
\draw(3.5,3.1) node{$x+z+(y \gg 9)+y$};

\draw[-stealth,color=blue] (2.0,2.95) to[out=200,in=120] (0.8,1.3);
\draw[-stealth,color=blue] (3.3,2.65) -- (3.15,1.3);
\draw[-stealth,color=blue] (3.8,2.65) -- (3.9,2.3);
\draw[-stealth,color=blue] (4.5,2.75) -- (5.2,2.3);

\draw[-stealth] (1.25,1.75) -- (1.1,1.3);
\draw[-stealth] (2.75,1.75) -- (2.9,1.3);
\draw[fill=white] (2,1.9) ellipse (10mm and 4.5mm);
\draw(2,2.05) node{$2\esp: +(0,1)$};
\draw(2,1.7) node{$x+y$};
\draw(1,1) ellipse (4mm and 3mm) node{$0\esp:x$};
\draw(3,1) ellipse (4mm and 3mm) node{$1\esp:y$};

\draw(4,2) ellipse (4mm and 3mm) node{$3\esp:z$};

\draw[-stealth] (5.75,1.75) -- (3.4,1.0); %
\draw[-stealth] (6.25,1.75) -- (6.5,1.3);
\draw[fill=white] (6,2) ellipse (10mm and 5mm);
\draw(6,2.15) node{$5\esp: \gg (1,4)$};
\draw(6,1.8) node{$y \gg 9$};
\draw(6.5,1) ellipse (4mm and 3mm) node{$4\esp:9$};

\setul{1pt}{.4pt}
\draw(2,2.55) node{\textbf{\ul{{Add:}}}};

\end{scope}


\end{tikzpicture}


  \caption{\label{egraph}Example of operations on an E-graph-style structure}
\end{wrapfigure}
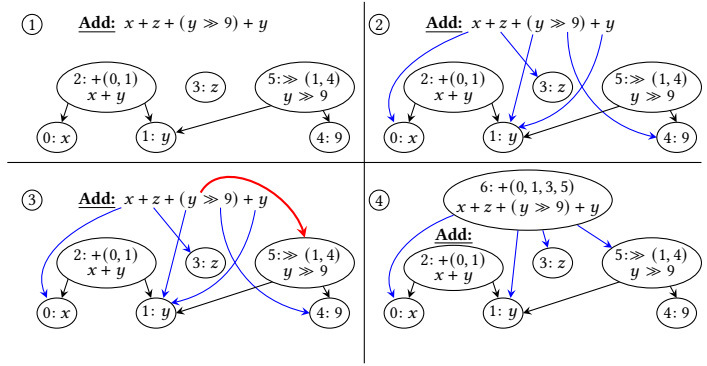
\cref{egraph} animates a simple example.
It steps through stages of adding a new node to the graph, representing the new symbolic expression: ($x + z + (y \gg 9) + y$).
Step 1 shows the initial state.
Nodes 2 and 5 are labeled with operators and IDs of operands.
To process the expression we are evaluating, we first look up existing graph nodes for all of its leaf expressions, as Step 2 shows.
Then we proceed bottom-up in the expression tree, finding an existing node or building a new one for each subexpression.
In this case, we next need to find a node for $y \gg 9$, in Step 3.
As we resolved $y$ to node 1 and $9$ to node 4, we are able to search the DAG for a node already labeled with operator $\gg$ and argument nodes 1 and 4, finding node 5.
In the final step, Step 4, we need to find a node for $x + z + (y \gg 9) + y$.
We have node IDs for all four operands of addition, and we \emph{sort} them by ID, taking advantage of associativity and commutativity of addition, to find a canonical description of this node.
No existing node has that description, so we add a new one.
The main complication absent from this example is normalization with rewrite rules going beyond associativity and commutativity; the approach is parameterized on a set of such rules (which must have proofs in Coq), and they are applied as each node of the input AST is processed.

\section{Related Work}\label{sec:related}

With \cryptopt we combine several known techniques to generate fast and formally verified code.
We now sketch related work in the areas of
genetic improvement (GI),
optimal optimization pass finding for off-the-shelf compilers,
superoptimization,
peephole optimization,
translation validation
and computer-aided cryptography.

\subsection{Genetic Improvement}

\cryptopt applies GI, an area within search-based software engineering~\cite{HARMAN2001833} that automatically searches out improved software versions. 
Genetic improvement is a relatively young research field; its first survey appeared in 2018~\cite{Petke2018gi}. 
Despite its youth, GI has already had real-world impact: maintainers have accepted GI patches into both open-source~\cite{Langdon2015:improving} and commercial~\cite{Haraldsson2017:fixing} projects. 

\cryptopt utilizes GI in the code-generation phase to generate fastest code per-microarchitecture.
To the best of our knowledge, \cryptopt is the first automatic compiler to offer both this level of microarchitecture-tuned performance (albeit for the limited domain of straight-line crypto code) and the highest level of formal assurance (Coq verification of all compiler phases that matter for soundness).
However, related work has tackled some of the constituent challenges.

\citet{BosamiyaGLPH20} is most similar to \cryptopt as they also use GI to find fast code per-architecture while being provably correct.
The primary objective is to parse optimized \xassembly and then use verified transformations to transform it back into a clean form, which is then easier to reason about.
From those transformations, they selected the ``prefetch insertion'' and ``instruction-reordering'' transformations and conducted a
case study on using GI to find fast implementations based on OpenSSL's AES-GCM (which uses AES-NI instructions).
Their approach can improve the performance of existing code and verify the correctness of the produced code.
Their starting point is handwritten assembly code within a relatively shallow metaprogramming framework. 
In contrast, we show that randomized search can be used as part of a fully automated pipeline that starts from high-level functional programs, allowing us to generate fast code for multiple elliptic curves, not just multiple target architectures from a single algorithm for Bosamiya et al.
Generating the code allows \cryptopt additional flexibility, with support for optimizing register allocation (in particular spills to memory) and instruction selection.
For example, using their reordering transformation, they cannot substitute an add-using-overflow~\instr{adox} for an add-using-carry~\instr{adcx}, let alone have those calculate two
independent additions in parallel, because they only consider reads and writes to the flag register in general, rather than the granularity of individual flags (i.e.\ read~\cf{}, write~\of{}
are independent).
\cryptopt also benefits from compatibility with \fiat, which makes code generation for finite fields for new primes easy.

\subsection{Optimization-Pass Finding}
\citet{StephensonOMA03} and \citet{PeelerLSRYB22} both use GI to select and order existing optimization passes of off-the-shelf compilers for optimal running time,
where the former uses a simulated running time (Trimaran \cite{Chakrapani2005}) as the objective function and the latter uses the actual running time.
\cryptopt, however, is not bound by either applying or not applying those fixed optimization passes from \ots compilers.
Rather, it explores many different variations for any particular code section and is also able to apply optimizations selectively at certain locations and avoid using them at others.

\subsection{Superoptimization}
\citet{Massalin87} coined the term ``superoptimizer'' to describe a tool for exhaustive enumeration of all possible programs to implement a given function.
The key idea making this feasible is the use of a probabilistic test set, which rejects the majority of incorrect candidates.
At the time of writing, it is able to generate programs of 12 instructions after several hours of running (on a 16MHz 68020 computer).
\citet{SasnauskasCCKTR17} present Souper, a tool to synthesize new optimizations on the LLVM IR.
Working on the IR, by design, they are unable to generate optimizations to exploit target-specific code sequences.

\citet{JoshiNR02} present Denali, a superoptimizer for very short programs.
They model the architecture of their processor (Alpha EV6) and use solvers to reject conjectures of the form ``No program can compute $P$ in at most 8 cycles.''
Combining this framing with a binary-search algorithm, they end up with the most efficient program.
\citet{Schkufza0A13} present STOKE, a superoptimizer which is able to synthesize and optimize programs.
It combines correctness indicator and performance into a cost function and then randomly (1) changes opcodes, (2) changes arguments, (3) deletes instructions, and (4) inserts
\instr{nop}s.
By starting from scratch, they can find algorithmically different solutions, which cannot be found by other superoptimizers.
The resulting programs range up to tens of instructions long.
Subsequent work extends STOKE to optimize floating-point kernels \cite{Schkufza0A14},
optimize loops \cite{SharmaSCA13},
and more aggressively optimize kernels based on verified runtime preconditions with cSTOKE \cite{SharmaSCA15}.

While \cryptopt shares the idea of a randomized search with superoptimization approaches, there are a few important differences between the two.
First, superoptimization approaches, and in particular STOKE, apply random changes to existing code, whereas \cryptopt aims to generate the code from a high-level specification.
Moreover, unlike superoptimization, \cryptopt only uses semantics-preserving approaches. 
This significantly reduces the search space, allowing \cryptopt to handle functions with hundreds and even thousands of instructions.
As shown in \cref{s:cvss}, superoptimizers tend to fail on the inputs that \cryptopt processes.
Finally, superoptimizers tend to use a model of the hardware for which they optimize the code.
\cryptopt, in contrast, optimizes to the actual hardware, allowing it to adapt to new hardware without the need to model the microarchitecture of the new hardware.

\subsection{Peephole Optimization}
Peephole optimizers use a sliding window on instructions (the peephole) and replace sets of instructions with more performant instructions~\cite{AhoSU86, COOPER2012597,
BERGMANN2003141}.
The replacement is usually done based on a predefined rule set (applying only to short instruction sequences), which itself is based on heuristics for estimating which set of
instructions is shorter or more performant.
Yet another approach is to find and learn good peephole optimizations automatically:
\citet{Bansal06} use machine-learning techniques to characterize small sections of code.
Then, based on those characteristics, they replace a code sequence with a semantically equivalent one assumed to be more performant.
While they only focus on small sections (on the order of tens of instructions),
\citet{PekhimenkoB10} use machine-learning techniques to characterize entire methods and then apply certain optimization transformations to them.
Similarly, \cryptopt considers the entire function as a whole, but rather than characterizing, learning and applying that knowledge to new functions,
\cryptopt considers each architecture and function as a black box and eventually finds a fast implementation.

\subsection{Verified Transformations}
Instead of proving the correctness of the compiler, translation validation~\cite{PnueliSS98} does not trust the compiler but verifies that the compiled code preserves the semantics of the source.
\citet{BosamiyaGLPH20}, as already mentioned, used their tool to transform optimized (manually written) \assembly code to easily verifiable code.
Similarly, TInA~\cite{Recoules19} lifts inline \assembly to semantically equivalent C code amenable to verification with known tools.
Only targeting the code for Curve25519, \citet{SchooldermanMSE21} used the Why3 proving platform~\cite{FilliatreP13} to verify an 8-bit AVR implementation. 

CryptoLine~\cite{Chen14,TsaiWY17,PolyakovTWY18,FuLSTWY19} is an automatic verification engine utilizing SMT solvers (BOOLECTOR) and computer-algebra systems (Singular), applying to their own IR.
The approach to validating assembly files is similar to ours.
CryptoLine has only worked via unverified translators from assembly languages to their IR, and the translator must be trusted, unlike ours, though it likely accepts some correct programs that ours rejects.

Last, \citet{SewellMK13} go further and parse the binary code of the {seL4}-Linux microkernel and transform it
until they could prove equivalence to the already-verified C code.

We would like to emphasize that those works aim to verify \emph{existing} implementations, whereas \cryptopt \emph{generates} them.
Targeting a wide range of microarchitectures for performance optimizations manually would quickly become practically infeasible.

\subsection{Real-World Applications of Computer-Aided Cryptography}
Provably correct generated code is already deployed in important projects: all major web browsers use finite-field code generated by \fiat~\cite{ErbsenPGSC19} (via Google's BoringSSL and other libraries), and Firefox includes routines arising from the more comprehensive efforts of Project Everest~\cite{Everest}, including compilation of nonstraightline code to C~\cite{Protzenko2017}, verified metaprogramming of assembly~\cite{FromherzGHPRS19}, tying it together in the EverCrypt library~\cite{ProtzenkoPFHPBB20}, and even adding protocol verification~\cite{BhargavanBK17}.
The Everest stack supports many different algorithms for the same functionality (e.g.\ AES+GCM or ChaCha20+Poly1305 for authenticated encryption)
and for each of those many different hand-optimized implementations depending on platform and hardware.
We already mentioned the work of \citet{BosamiyaGLPH20} on automatic program search, the only approach to \emph{automatic} assembly generation that we have seen in the Everest ecosystem, and it does not seem to have been applied yet to elliptic curves.
\cryptopt also has the usual advantage of proof-assistant work, that, despite our usage of a stack of domain-specific tools, none of them need be trusted.

\citet{BelyavskyBCRU20} published work on generating prime-agnostic point arithmetic in C using verified field arithmetic
from \fiat and claim timing-side-channel-resistant code; however, there is no formal verification of correctness or constant time.

\section{\cryptopt Overview}\label{s:oursolution}
Our aim is to strengthen \fiat to both increase the performance of the produced code and to decrease the size of the trusted code base.
To that end, we implement two novel components and integrate them with 
\fiat as sketched in \cref{fig:contrib}.

\begin{wrapfigure}{r}{0.3\textwidth}
    \centering
    \tikzset{every picture/.style={line width=0.75pt}} 
\begin{tikzpicture}[-latex, yscale=.7,xscale=.7]
    \scriptsize
    \tikzstyle{codeblock} = [draw=black, inner sep=.4em, text width=5.5em, minimum height=3em, align=center];

    \node[codeblock] (in) at (7,7) {Generate Initial Code};

    \node[codeblock] (code) [below=1.5em of in] {Mutate};
    \draw[-latex] (in.south) -- (code.north){};

    \node[codeblock] (measure) [below=1.5em of code] {Measure};
    \draw[-latex] (code.south) -- (measure.north){};

    \node[diamond, draw, below=1.5em of measure, text width=5em, align=center] (worse) {Worse\\Performance?};
    \draw[-latex] (measure.south) -- (worse.north){};

    \draw[-latex] (worse.east) -- ++(1,0) node[yshift=.6em,midway] {No} |- (code.east){};

    \node[codeblock] (revert) [below=1.5em of worse] {Revert Mutation};
    \draw[-latex] (worse.south) --  node[xshift=0.9em] {Yes}  (revert.north){};
    \draw[-latex] (revert.south) |- ++(-2,-1.5em) |-  (code.west){};

\end{tikzpicture}
    \caption{Optimizer system architecture\label{fig:concept} }
\vspace{-2mm}
\end{wrapfigure}
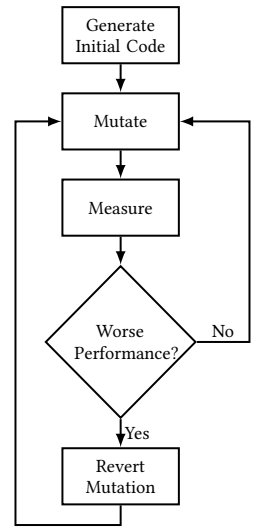

The first component, the \cryptopt optimizer, is a new backend for \fiat, which ingests \fiatir and produces \xassembly code.
A unique and novel property of the optimizer is that instead of relying on classic compiler-optimization techniques, it draws on techniques from the domain of evolutionary computation.
Specifically, as illustrated in \cref{fig:concept}, the \cryptopt optimizer first randomly generates \xassembly code that implements the input function.
It then repeatedly mutates the code and measures the execution time of both the original and the mutated code, discarding the slower one.
The process continues until a predefined computational budget is used up.

The second component is a new program-equivalence checker, which, given an original \fiatir program and its optimized assembly, is able to verify behavior preservation with no
further hints (c.f. \cref{fig:contrib}).
It is codesigned with the \cryptopt optimizer to support the same set of transformations, significantly reducing the complexity compared to a generic equivalence checker.
The checker itself is implemented and verified in Coq.
Thus, only \fiat and the checker are verified, whereas the optimizer itself need not be trusted.
While not trusted, the optimizer is designed to only use semantics-preserving transforms. 
Thus, during normal operation, we expect verification always to succeed.

The combination of these two components allows us to achieve our aims.
As we demonstrate, random search can generate code that performs as well as hand-optimized code, significantly surpassing the performance of compiler-produced code.
At the same time, the formally verified checker strengthens the unified theorems of compilation to cover the whole span from functional programs to \xassembly instead of \fiatir, removing the compiler from the trusted code base.

While designed as a backend for \fiat, \cryptopt can operate as a stand-alone optimizer. 
To demonstrate this use case, we transform a C implementation of field arithmetic to LLVM IR using Clang. (Specifically, we use the C implementation of the Bitcoin \texttt{libsecp256k1} library.)
We then use a simple script to translate the LLVM IR to the input format of \cryptopt and optimize it.
Because the C code we use is not formally verified, we skip the formal-verification aspect in this use case, focusing just on further evaluation of the randomized optimizer.

Finally, we note that the generated code from \fiat utilizes neither secret-dependent memory accesses nor secret-dependent branching.
Consequently, the code follows the constant-time programming paradigm~\cite{AlmeidaBBDE16}, providing protection against microarchitectural side-channel attacks~\cite{GeYCH18}.

Now we are ready to fill in the details of \cryptopt (randomized search in \cref{s:randomsearch} and equivalence checking in \cref{s:equivalence}) and how we evaluated it (\cref{s:evaluation}).

\section{Randomized Search for Assembly Programs}\label{s:randomsearch}

The first half of our approach is the \cryptopt optimizer, which generates highly performant \xassembly code that  implements an input \fiatir function.
A unique feature of \cryptopt is that instead of relying on heuristics for generating the code, \cryptopt explicitly casts the problem as a combinatorial optimization problem.
That is, we observe that searching the set of assembly code sequences that implement a given input function is discrete.
Hence, the problem of searching this set for the assembly code sequence that minimizes execution time is a combinatorial optimization problem.

For optimization, we employ the \rls strategy with the \barheur heuristics.
Recall that for \rls, we need to first choose a random solution and then repeatedly mutate the solution.
While \rls is a relatively simple approach for combinatorial optimization, we find that it is effective and, as we show below, achieves good results.
We leave the task of experimenting with more complex optimization strategies to future work.
In the rest of this section we present our approach for generating and mutating solutions.
We start with a description of the input format and then explain how we generate and mutate code.

\subsection{Input Format}

\begin{wrapfigure}{r}{0.5\textwidth}
    \vspace{-1em}
\begin{minipage}{\linewidth}
  \scriptsize
  $$\begin{array}{rrcl}
    \textrm{Variable} & x \\
    \textrm{Binary integer} & b \\
    \textrm{Operand} & e &::=& x \mid b \\
    \textrm{Operator} & o &::=& ! \mid \& \mid * \mid + \mid - \mid \; << \; \mid \; = \; \mid \; >> \; \mid \; \sim \; \mid \\
      &&&  \mathsf{or} \mid \mathsf{addcarryx} \mid \mathsf{cmovznz} \mid \mathsf{mulx} \mid \\
      &&&  \mathsf{static\_cast} \mid \mathsf{subborrowx} \\
    \textrm{Expression} & E &::=& \mathsf{return} \; e \mid x, \ldots, x \leftarrow o(e, \ldots, e); E
  \end{array}$$
\vspace{-3mm}
  \caption{\fiatir syntax\label{sourcesyntax}}
  \end{minipage}
  \vspace{-3mm}
\end{wrapfigure}

Recall (\cref{s:oursolution}) that \cryptopt takes \fiatir as an input.
This intermediate language is truly a minimal one (see \autoref{sourcesyntax}), with the only noteworthy syntactic twist being that integer constants are presented as binary numbers (clearly indicating bitwidth), though we will often abbreviate them in decimal form, when bitwidth is clear from context.
Operators may in general take not just multiple operands but also generate multiple results, associated with less common operators like addition with carry or multiplication producing double-wide results via two output words.

Input programs contain no branches or explicit memory accesses, just accesses of local variables.
As a result, generated assembly programs will not be too much more complex, avoiding all memory aliasing and restricting pointer expressions to be constant offsets of either function parameters (for data structures passed by reference) or the stack pointer (for spilled variables).
The \assembly we generate is timing-secure for the same reasons that \fiatir is timing-secure;
we only use primitive program mutations that either preserve relevant behavior (the trace of program memory
accesses and control-flow decisions) or add new behaviors in ways independent of secrets (all memory accesses are to constant offsets from either the stack pointer or arrays
that are function parameters), so by induction the initially secure programs are still secure after optimization.

\newcommand{\ali}[2]{\makebox[#1][r]{#2}}
\begin{wrapfigure}{r}{0.63\textwidth}
    \vspace{-1em}
    \begin{minipage}{0.31\textwidth}
        \begin{algorithm}[H]
            \scriptsize
            \SetKwInOut{Input}{input}\SetKwInOut{Output}{output}
            \SetKwProg{Function}{function}{ \\begin}{end}
            \SetKwBlock{Block}{\{}{\}}
            \Input{$X, Y, Z \textrm{ such that } 0\leq X, Y, Z < 2^{63}$}
            \Output{$O=2^{64}O_1+O_0=Z^2 + (Y+Z) \cdot X + Z$}
            \BlankLine
            \Function{example($X,\ Y,\ Z$)}{
                \ali{1em}{$t_2,      $} \ali{1.4em}{$ t_1 $} $\leftarrow \fmul_1(Z,  $ \ali{1.4em}{$Z)$}\\
                \ali{1em}{$\cnothing,$} \ali{1.4em}{$ t_0 $} $\leftarrow \fadd_1(Y,  $ \ali{1.4em}{$Z, $}  \ali{1.4em}{$0  )$}\\
                \ali{1em}{$t_4,      $} \ali{1.4em}{$ t_3 $} $\leftarrow \fmul_2(t_0,$ \ali{1.4em}{$X)$}\\
                \ali{1em}{$c_0,      $} \ali{1.4em}{$ t_5 $} $\leftarrow \fadd_2(t_3,$ \ali{1.4em}{$t_1,$} \ali{1.4em}{$0  )$}\\
                \ali{1em}{$c_1,      $} \ali{1.4em}{$ O_0 $} $\leftarrow \fadd_3(t_5,$ \ali{1.4em}{$Z  ,$} \ali{1.4em}{$0  )$}\\
                \ali{1em}{$\cnothing,$} \ali{1.4em}{$ t_6 $} $\leftarrow \fadd_4(t_4,$ \ali{1.4em}{$t_2,$} \ali{1.4em}{$c_0)$}\\
                \ali{1em}{$\cnothing,$} \ali{1.4em}{$ O_1 $} $\leftarrow \fadd_5(t_6,$ \ali{1.4em}{$0  ,$} \ali{1.4em}{$c_1)$}\\
                \Return{$O_1,\ O_0$}
            }
            \caption{An Example Function\label{alg:example}}
        \end{algorithm}
    \end{minipage}\hfill%
    \begin{minipage}{0.3\textwidth}
        \centering
            \begin{tikzpicture}[-latex,scale=0.7,font=\scriptsize]
    \tikzstyle{operation} = [rounded rectangle, draw=black, minimum width=1cm];

    \node (X) at (0.6,-0.6) {$X$};
    \node (Y) at (2.0,-0.6) {$Y$};
    \node (Z) at (4.0,-0.6) {$Z$};

    \node [operation, \colA] (A1) at (2.4, -1.5) {$\fadd_1$};
    \node [operation, \colB] (M1) at (4.0, -1.5) {$\fmul_1$};
    \node [operation, \colC] (M2) at (1.0, -2.5) {$\fmul_2$};

    \node [operation, \colE] (A2) at (4, -3.4) {$\fadd_2$};
    \node [operation, \colD] (A3) at (4, -4.5) {$\fadd_3$};

    \node [operation, \colF] (A4) at (1, -4.5) {$\fadd_4$};

    \node [operation, \colG] (A5) at (1, -5.5) {$\fadd_5$};

    \node (O1) at (1, -6.5) {$O_1$};
    \node (O0) at (4, -6.5) {$O_0$};

    \draw[color=black!50] ($(Y.south)-(0,0)$)  -- ($(A1.north)-(.4,0)$);
    \draw[color=black!50] (Z)                  -- ($(A1.north)+(.4,0)$);

    \draw[color=black!50] ($(Z.south)-(.1,0)$) -- ($(M1.north)-(.1,0)$);
    \draw[color=black!50] ($(Z.south)+(.1,0)$) -- ($(M1.north)+(.1,0)$);

    \draw[color=black!50] (X)                  --                             ($(M2.north)-(.4,0)$);
    \draw                 (A1.south)           -- node [pos=0, below] {$t_0$} ($(M2.north)+(.4,0)$);

    \draw ($(M2.south)+(.4,0)$) -- node [pos=0, below,       inner sep=1] {$t_3$} ($(A2.north)-(.4,0)$);
    \draw (M1.south)            -- node [pos=0, below right, inner sep=1] {$t_1$} (A2.north);

    \draw ($(M2.south)-(.4,0)$) -- node [pos=0, below left, inner sep=1] {$t_4$} ($(A4.north)-(.4,0)$);
    \draw ($(M1.south)-(.4,0)$) -- node [pos=0, below] {$t_2$}                   (A4.north);
    \draw ($(A2.south)-(.4,0)$) -- node [pos=0, below] {$c_0$}                   ($(A4.north)+(.4,0)$);

    \draw                  (A2.south)          -- node [pos=0,  below right, inner sep=1] {$t_5$} (A3.north);
    \draw[color=black!50, bend left] ($(Z.south)+(.2,0)$) to[out=50,in=130] ($(A3.north)+(.4,0)$);

    \draw ($(A3.south)-(.4,0)$) -- node [pos=0, below] {$c_1$}      ($(A5.north)+(.4,0)$);
    \draw ($(A4.south)-(.4,0)$) -- node [pos=0, below left] {$t_6$} ($(A5.north)-(.4,0)$);

    \draw (A3.south) -- (O0);
    \draw ($(A5.south)$) -- (O1);

\end{tikzpicture}
        \captionof{figure}{Data flow of the running example in \cref{alg:example}\label{fig:dataflow}}
     \end{minipage}
\end{wrapfigure}

\cref{alg:example} shows a \fiatir program, which we will use throughout this paper as a running example.
The program takes three inputs ($X, Y, Z$) and outputs $Z^2 + (Y+Z) \cdot X + Z$ using two types of operations: \fadd and \fmul.
The operation \fadd adds two 64-bit numbers and one 1-bit carry, then returns the sum as one 64-bit number and one 1-bit carry.
The operation \fmul multiplies two 64-bit numbers, then returns the 128-bit product as two 64-bit words.
For simplicity, we assume that the arguments are in the range $0\leq X,\ Y,\ Z < 2^{63}$,
allowing us to ignore some carries known to be~0. 
We mark these carries with $\cnothing$.

In the absence of memory aliasing and control flow, the restrictions on operation evaluation order in the programs we optimize are captured by the data flow.
\cref{fig:dataflow} shows the data-flow graph of the code in \cref{alg:example}.

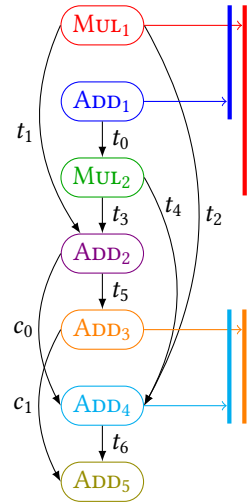
\begin{wrapfigure}{r}{.29\linewidth}
    \vspace{-1em}
    \begin{tikzpicture}[-latex]
    \tikzstyle{operation} = [rounded rectangle, draw=black, minimum width=1cm]
    \tikzstyle{line} = [-, ultra thick]

    \node [operation, \colB] (M1) at (0, -0) {$\fmul_1$};
    \node [operation, \colA] (A1) at (0, -1) {$\fadd_1$};
    \node [operation, \colC] (M2) at (0, -2) {$\fmul_2$};
    \node [operation, \colE] (A2) at (0, -3) {$\fadd_2$};
    \node [operation, \colD] (A3) at (0, -4) {$\fadd_3$};
    \node [operation, \colF] (A4) at (0, -5) {$\fadd_4$};
    \node [operation, \colG] (A5) at (0, -6) {$\fadd_5$};

    \draw             (A1.south) to node[right]{$t_0$} (M2.north);

    \draw[bend right] (M1.west) to node[midway, xshift=-.2cm]{$t_1$}         (A2.north west);
    \draw[bend left]  (M1.east) to node[midway, xshift= .2cm]{$t_2$}         (A4.east);

    \draw             (M2.south) to node[right]{$t_3$}                       (A2.north);
    \draw[bend left]  (M2.east)  to node[very near start, xshift=.2cm]{$t_4$}(A4.east);

    \draw             (A2.south) to node[right]{$t_5$}                       (A3.north);
    \draw[bend right] (A2.west)  to node[midway, xshift=-.2cm]{$c_0$}        (A4.west);

    \draw             (A4.south) to node[right]{$t_6$}                       (A5.north);
    \draw[bend right] (A3.west)  to node[midway, xshift=-.2cm]{$c_1$}        (A5.west);
    \draw [line, ->, thin, \colB] (M1.east) -> ++(+1.3,0);
    \draw [line, \colB] (M1.north east) +(+1.6,0) -- ++(+1.6,-2.5);
    \draw [line, ->,  thin, \colA] (A1.east) -- ++(+1.1,0);
    \draw [line, \colA] (M1.north east) +(+1.4,0) -- ++(+1.4,-1.5);
    \draw [line, ->, thin, \colD] (A3.east) -- ++(+1.3,0);
    \draw [line, \colD] (A3.north east) +(+1.6,0) -- ++(+1.6,-1.5);
    \draw [line, ->, thin, \colF] (A4.east) -- ++(+1.1,0);
    \draw [line, \colF] (A3.north east) +(+1.4,0) -- ++(+1.4,-1.5);

\end{tikzpicture}
    \vspace{-.5em}

    \caption{One ordering. Round rectangles show the operations, which are evaluated top-down. Arrows indicate creation and consumption of intermediate values. Vertical bars show
    the intervals in which an operation can be scheduled.} \label{fig:oneordering}
    \vspace{-1em}
\end{wrapfigure}

\subsection{Code Generation}

The process of generating those assembly candidates can be understood as split between instruction scheduling, instruction selection, and register allocation, in that order.
Each phase makes certain arbitrary decisions that may be changed by a later random mutation.
We summarize the phases here before returning to details of generation and mutation.

First, for instruction scheduling (see \cref{s:instrSched}), we use data-flow analysis to determine the data-flow dependencies between operations and choose a random topological order of the
dependency graph as the initial order of operations in the code.
Then, for instruction selection (see \cref{s:instrSelec}), each operation gets assigned a compatible \xassembly instruction template.
Finally, in our setting, register allocation (see \cref{s:registerAllocation}) arises mostly in ensuring compatibility with operand restrictions of the instruction templates that were selected.
For example, consider the objective to \emph{multiply two values} (one single instruction can at most read from one memory location), in a context where both operands reside in memory.
The relevant dimension of freedom is which value to load into a register explicitly and into which register.
After the decisions of the three phases have been recorded, it is easy to read off the chosen assembly program sequentially.
Recall that all of these decisions may be revisited later in random mutations.

To present details of the three phases, we focus on the example program from \cref{alg:example}.
\cref{fig:oneordering} shows how operations are initially ordered and potential scheduling intervals;
the mutations are shown in \cref{fig:mutations};
until finally we see the emitted code with and without the effects of the mutations in \cref{lst:emitedCode}.
We will reference those illustrations later as needed.

\subsection{Instruction Scheduling}
\label{s:instrSched}
\begin{wrapfigure}{r}{.35\textwidth}\vspace{-4mm}
        \begin{tikzpicture}[-latex]
        \tikzstyle{operation} = [rounded rectangle, draw=black, minimum width=1cm]
        \tikzstyle{instr} = [align=center,baseline, ellipse, minimum height=1.2em,  draw=black, fill=white, text width=1.8em, inner sep=0em, xshift=-.6em, yshift=-.7em, font={\footnotesize}]

        \node (I)   at (0,-.25) {$\mathit{I}$};
        \node (II)  at (2.5,-.25) {$\mathit{II}$};

        \node [operation, \colB] (M1-1) at (0, -1) {$\fmul_1$};
        \node [operation, \colA] (A1-1) at (0, -2) {$\fadd_1$};
        \node [operation, \colC] (M2-1) at (0, -3) {$\fmul_2$};
        \node [operation, \colE] (A2-1) at (0, -4) {$\fadd_2$};
        \node [operation, \colD] (A3-1) at (0, -5) {$\fadd_3$};
        \node [operation, \colF] (A4-1) at (0, -6) {$\fadd_4$};
        \node [operation, \colG] (A5-1) at (0, -7) {$\fadd_5$};

        \node [instr, right of=A1-1] (iA1-1) {\instr{adcx}};
        \node [instr, right of=M2-1] (iM2-1) {\instr{mulx}};
        \node [instr, right of=M1-1] (iM1-1) {\instr{mulx}};
        \node [instr, right of=A2-1] (iA2-1) {\instr{add}};
        \node [instr, right of=A3-1] (iA3-1) {\instr{add}};
        \node [instr, right of=A4-1] (iA4-1) {\instr{add}};
        \node [instr, right of=A5-1] (iA5-1) {\instr{add}};

        \node [operation, \colB] (M1-2) at (2.5, -1) {$\fmul_1$};
        \node [operation, \colA] (A1-2) at (2.5, -2) {$\fadd_1$};
        \node [operation, \colC] (M2-2) at (2.5, -3) {$\fmul_2$};
        \node [operation, \colE] (A2-2) at (2.5, -4) {$\fadd_2$};
        \node [operation, \colF] (A4-2) at (2.5, -5) {$\fadd_4$};
        \node [operation, \colD] (A3-2) at (2.5, -6) {$\fadd_3$};
        \node [operation, \colG] (A5-2) at (2.5, -7) {$\fadd_5$};

        \node [instr, right of=M2-2] (iM2-2) {\instr{mulx}};
        \node [instr, right of=A1-2] (iA1-2) {\instr{add}};
        \node [instr, right of=M1-2] (iM1-2) {\instr{mulx}};
        \node [instr, right of=A2-2] (iA2-2) {\instr{add}};
        \node [instr, right of=A3-2] (iA3-2) {\instr{add}};
        \node [instr, right of=A4-2] (iA4-2) {\instr{add}};
        \node [instr, right of=A5-2] (iA5-2) {\instr{add}};

        \draw [bend left=55, dashed] (iA1-1.north) to node[near start, xshift=.4cm, yshift=.3cm]{\footnotesize$\beta$} (iA1-2.north);
        \draw [dashed] (A4-1.north east) -- node[midway, xshift=.4cm, yshift=-.05cm]{\footnotesize$\alpha$} (A4-2.south west);

    \end{tikzpicture}
    \centering
    \vspace{1em}
    \caption{$\mathit{I}$: Initial ordering of operations (colored rounded rectangles) with attached templates (black ellipses); $\mathit{II}$: After two mutations $\alpha$ and $\beta$:
        mutation $\alpha$ in topological ordering, mutation $\beta$ in instruction-template selection. Data-flow arrows omitted; dashed arrows indicate mutations.
    \label{fig:mutations}}
    \vspace{-5mm}
\end{wrapfigure}
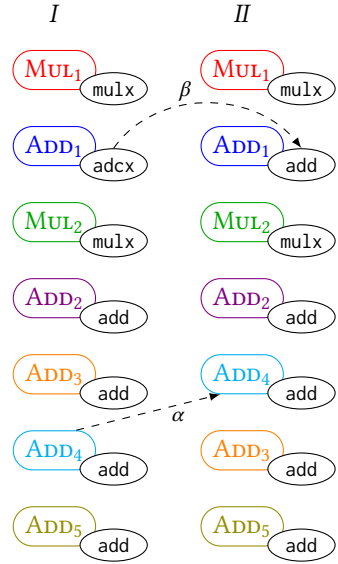
Thanks to the simplicity of \fiatir, every variable is assigned exactly once in the straight-line programs that the \cryptopt optimizer takes as input.
Consequently, any operation can be evaluated whenever all its inputs have been computed.
In the example, $\fmul_2$ can be evaluated when~$t_0$ and~$X$ are computed. 

\parhead{Initial Ordering.}
To create an initial ordering, the \cryptopt optimizer simply computes a topological order of the data-flow graph.
\cref{fig:oneordering} shows an example of an initial ordering for our running example.
We would like to emphasize that the selection does not rely on any heuristic.
While this initial selection does affect the subsequent mutated ordering,
our mutation strategy guarantees that any possible ordering can be reached
from any initial random starting point via a sequence of mutation steps.

\parhead{Mutation Step.}
\newcommand{\op}{$\mathit{op}$\xspace}
An instruction-scheduling mutation step randomly selects one operation in the current ordering.
This operation is moved to a randomly chosen location within the interval where it would be valid to move: not before the last assignment responsible for setting a variable that is used here as an operand, nor after the first assignment that reads the variable being set here.
We begin with the ordering shown in \cref{fig:oneordering}.
The vertical colored bars indicate the intervals where the respective operations can be scheduled.
Step~$\alpha$ of \cref{fig:mutations} shows the effect of moving $\fadd_4$ up one position.

Selecting the position to move an operation to is biased towards larger moves, i.e.\ further away from the initial position, in an effort to minimize spills by minimizing distance between computing a value and using it.

\subsection{Instruction Selection}
\label{s:instrSelec}

An important property of complex instruction sets such as x86-64 is that there can be multiple alternative implementations for each high-level operation, each with slightly different semantics and impact on the processor pipeline.
To match  machine instructions to operations,
the \cryptopt optimizer uses templates describing the possible implementations for each operation.

Consider, for example, the \fadd operation, which can be implemented using multiple instructions, such as \instr{add}, \instr{adcx}, or \instr{adc}.
Though semantically equivalent, these choices influence program state differently.
For instance, in the case of no input carry ($\fadd_{1-3}$), implementing an \fadd operation using an \instr{adcx} instruction requires clearing the carry flag.
The template of the \instr{adcx} instruction accounts for that and issues a clear-carry instruction (\instr{clc}) before the \instr{adcx} instruction, unless \cryptopt determines that the carry is already clear.
Lines~6 and~7 of \cref{lst:emitedCode} show this choice for $\fadd_1$.

\parhead{Initial Template Mapping.}
When creating the initial code, \cryptopt selects a random template for each operation as an initial mapping.
\cref{fig:mutations} ($\mathit{I}$) shows an example of a possible mapping of templates to the operations of our example function.
The figure indicates the operation $\fmul_1$ is implemented using the \instr{mulx}, $\fadd_1$ uses \instr{adcx}, and so forth.

\parhead{Mutation Step.}
To mutate the instruction selection, \cryptopt chooses an operation at random and replaces the template for that instruction with one alternative.
Mutation $\beta$ in \cref{fig:mutations} shows an example of replacing the template for $\fadd_1$ to use the \instr{add} instruction instead of the original \instr{adcx}.

\parhead{Flag Spills.}
The series of additions $\fadd_2, \dots, \fadd_5$ show not only the influence of different orderings on the resulting code
but also how a template is used to handle flag spills.
Consider the code in \cref{lst:emitA} where $\fadd_2$ results in \cf{}=$c_0$.
Later on, $\fadd_3$ needs to write its own \cf{}=$c_1$.
Therefore, the current \cf{} needs to be spilled (into another register).
In this case, it does so with \instr{setc dl} (line~17).
Similarly, $\fadd_4$ then needs to spill $c_1$ (line~22) to avoid overwriting it with its $\cnothing$.

\begin{wrapfigure}{r}{.5\textwidth}
    \scriptsize
    \vspace{-1em}
    \input{figures/tex/assembly}
    \vspace{-1em}
    \caption{Emitted assembly code. Highlighted lines show effects from mutations.}
    \label{lst:emitedCode}
    \vspace{-3em}
\end{wrapfigure}

At this point (line~23), \cryptopt needs to add the values of three operands,
i.e.\ $t_4 + t_2 + c_0$.
As the \xassembly language does not have a single three-operand addition instruction,
\cryptopt first adds two operands together then adds the sum to the third one.
Note that this changes the evaluation order from $t_4 + t_2 + c_0$ to $(t_2 + c_0) + t_4$.
As the \fadd operation is associative and commutative, any evaluation order maintains correctness.
The equivalence checker accounts for this change in evaluation order (see \cref{s:equivalenceEGraph}).

\parhead{Strength Reduction.}
Some of the templates we use support limited forms of strength reduction.
For example, we have templates to implement multiplication by a constant, including using a left-shift operation (i.e.\ $x \times 8 \implies x \ll 3$), a series of
multiplications and additions (i.e.\ $x \times 5 \implies x \times 2 + x$), or a combination thereof.
The final template selection is left to the optimizer.

\subsection{Register Allocation}\label{s:registerAllocation}

The register-allocation step of \cryptopt achieves two aims.
It must both decide which values are assigned to registers and which registers to spill to memory when running out of registers.
\cryptopt uses randomized search for the former and a deterministic strategy for the latter.

\parhead{Register Assignment.}
Registers need to be assigned in two main cases: when computing a new value and when reading a value from memory, either from the input or following a register spill.
\cryptopt keeps track of the live registers, allocating a free register if one is available.
If none is available, \cryptopt spills the contents of a register to memory and uses the freed register.
To choose the register to spill, \cryptopt scans the future use of all registers and spills the register whose next use is furthest based on the current operation order.

For example, lines~10--11 in \cref{lst:emitedCode} implement the $\fmul_2$ operation.
(For this example, we assume that the architecture only has three general-purpose registers: \instr{r8}, \instr{r9}, and \instr{rdx}.)
At this point, registers \instr{r8}, \instr{r9}, \instr{rdx} have been used for $t_2$, $t_1$, and $t_0$, respectively.
Hence, the need to spill a register.
Observing that $t_2$ is not required until $\fadd_4$, whereas $t_0$ and $t_1$ are used earlier,
\cryptopt spills \instr{r8} (line~10).

\parhead{Memory Loads.}
Most arithmetic operations in the x86-64 architecture support instruction formats that take one argument from memory.
When an argument of an operation is in memory, \cryptopt tries to use such an instruction format.
When this is not possible, e.g.\ when the values of two arguments are in memory, \cryptopt resorts to loading a value from memory into a register.
In the case of associative operations, such as addition and multiplication, \cryptopt initially randomly chooses the value to load, though mutations may later alter the choice.

\parhead{Exploiting Simplicity.}
Finally, we note that the absence of control flow and avoidance of human heuristics are key enablers for memory-spill decisions.
The former simplifies dependency analysis, allowing \cryptopt to determine the requirements for downstream operations.
The latter allows \cryptopt to examine the entire function
rather than focusing on instructions within a peephole window, a technique commonly used by \ots compilers.

\subsection{Objective-Function Evaluation}
\label{sec:costfunctionevaluation}
We compare different random program variants by running them on the actual processors of interest.
Controlling noise in running-time measurement is of utmost importance because with too much noise, randomized search could be driven into unproductive oscillation.
We found the details of such measurement surprisingly difficult to get right.
\full{\cref{app:costfunctionevaluation}}{The full version of this paper} gives those details, which rely on running a chosen number of repetitions of the two candidate programs, interleaved in a random order, then returning the median timing observed per program.

\section{Checking Program Equivalence}\label{s:equivalence}

\begin{figure}
  \footnotesize \begin{lstlisting}
Inductive REG := rax | rcx | (* ...65 omitted... *) | r15b.
Inductive AccessSize := byte | word | dword | qword.
Record MEM := { mem_bits_access_size : option AccessSize; mem_base_reg : option REG;
                mem_scale_reg : option (Z * REG); mem_base_label : option string; mem_offset : option Z }.
Inductive FLAG := CF | PF | AF | ZF | SF | OF.
Inductive OpPrefix := rep | repz | repnz.
Inductive OpCode := adc | adcx | add | adox | and | bzhi | call | clc | cmovb | cmovc | cmovnz | cmp
  | db | dd | dec | dq | dw | imul | inc | je | jmp | lea | mov | movzx | mul | mulx | pop | push
  | rcr | ret | sar | sbb | setc | seto | shl | shlx | shr | shrx | shrd | sub | test | xchg | xor.
Record JUMP_LABEL := { jump_near : bool; label_name : string }.
Inductive ARG := reg (r : REG) | mem (m : MEM) | const (c : Z) | label (l : JUMP_LABEL).
Record NormalInstruction := { prefix : option OpPrefix; op : OpCode; args : list ARG }.
  \end{lstlisting}
\vspace{-3mm}
  \caption{Syntax of Coq embedding of \xassembly\label{destsyntax}}
\end{figure}

\noindent
Our goal with \cryptopt was to preserve or even strengthen the formal guarantees of \fiat.
One way to achieve that goal would have been to verify the whole randomized-search process with Coq, but we wanted to find a simpler strategy that would have the side benefit of also potentially supporting automatic verification of various handwritten assembly solutions.
Therefore, we decided to write a program-equivalence checker in Coq and verify it.
Industrial-strength translation validation as in Alive~\cite{Alive2} is now well-established, but again, proving such a tool from first principles would be a substantial undertaking.
We were curious, instead, how far we could get implementing (and proving) our checker from scratch, lifting just those features of more conventional checkers that turned out to be important in our domain.

\autoref{destsyntax} shows a nearly complete description of the \xassembly syntax accepted by our checker.
For better error messages, some control-flow opcodes like \texttt{jmp} are included, though they will always be rejected by the checker.
This type of syntax trees is given a very standard semantics, in the form of an interpreter as a Coq function.
The simplicity of syntax and semantics is important, since both are referenced by the final theorem for any specific compilation, while the syntax and semantics of \fiatir drop out of the picture as untrusted.

\subsection{Code Verification}
The best-performing implementation produced by \cryptopt is assured to be correct through a formally verified equivalence checker.
Specifically, we verify the correctness of \cryptopt's output through functional equivalence between programs in the \fiatir and \xassembly.
We developed simple symbolic-execution engines for the relevant subsets of both languages, producing program-state descriptions in a common logical format.
The next task is to check that function-output registers and memory locations store provably equivalent values between the two programs.
To that end, we developed a simple equivalence theorem prover that borrows from SMT solvers,
using a similar (E-graph) data structure.

The public API connecting the checker's two main components is based on the following definition of expressions:
$$\begin{array}{rrcl}
	\textrm{Integer constants} & n \\
	\textrm{Variables} & x \\
	\textrm{Operators} & o \\
	\textrm{Expressions} & e &::=& n \mid x \mid o(e, \ldots, e)
\end{array}$$
The E-graph exposes a function \texttt{internalize} that takes in an expression and returns a variable now associated with that expression's value.
Importantly, the expression will usually mention variables that came out of previous calls, which lets us work with exponentially more compact representations than if we expanded out all variables.
The internal E-graph takes advantage of this sharing for efficiency.
Also, crucially, every \texttt{internalize} invocation proactively infers equalities between previously considered expressions and the new expression and its subterms.
Thus, we may check two expressions for equality simply by verifying that \texttt{internalize} maps them to the same variable, which becomes a chosen representative of an equivalence class of expressions.

\subsection{Symbolic Execution}

We built symbolic-execution engines for the two relevant languages, \fiatir and \xassembly.

The engine for the IR is simpler than for \xassembly.
Programs in this IR are just purely functional sequences of variable assignments with expressions that effectively already match the grammar we just gave.
Thus, to evaluate such a program symbolically, we just maintain a dictionary associating program variables to logical variables; the latter are effectively handles into the E-graph.

The execution engine for \xassembly is moderately more complicated.
Now the symbolic state associates not program variables but \emph{registers and memory addresses} with logical variables.
We take advantage of the stylized structure of cryptographic code to simplify the treatment of memory.
The only valid pointer expressions are constant offsets from either function parameters (standing for cells within data structures passed to the function) or the stack pointer (standing for spilled temporaries).
Thus, it is appropriate to make the symbolic memory a dictionary keyed off of \emph{pairs of logical variables and integers}.
The logical variable is the base address of an array in memory, while the integer gives a fixed offset into one of its words.

With this convention fixed, it is fairly straightforward to march through the instructions in a program, updating the register and memory dictionaries with the results of \texttt{internalize} calls.
The symbolic executor effectively breaks each (possibly complex) x86 instruction down into multiple simpler operations on integers.
There are multiple operations because many instructions affect both flags and their explicit destination registers.
The explanations of those effects are expressions from the grammar above, using a relatively small vocabulary of bitvector operators.

At the end of symbolic execution of an \assembly function, we pull the output values out of calling-convention-designated registers and memory locations.
These can then be compared against the explicit return value of a \fiatir program.
Both are expressed as logical variables connected to a common E-graph, so they should be syntactically equal exactly when the E-graph found a proof of equality.
Importantly, both symbolic states are initialized with common logical variables.

\subsection{Equivalence}

\label{s:equivalenceEGraph}
\begin{wrapfigure}{r}{0.5\textwidth}
    \vspace{-1em}
\begin{minipage}{\linewidth}
\begin{algorithm}[H]
  \scriptsize
	\SetKwInOut{Input}{input}\SetKwInOut{Output}{output}
	\SetKwProg{Function}{function}{ \\begin}{end}
	\SetKwBlock{Block}{\{}{\}}
	\Input{$\mathit{Op}$ an operator, $\mathit{Args}$ its list of argument variables}
	\Output{$n$, a variable / graph node for the expression's equivalence class}
	\BlankLine
	\Function{internalize($\mathit{Op},\mathit{Args}$)}{
		\If{Op is associative}{ 
			\For{argument $a$ in $\textit{Args}$}{
				\If{$a$ is also labeled with Op}{
					Expand $a$ in the list into its own E-graph neighbors
				}
			}
		}
		\If{Op has identity element e}{ 
			Remove from $\mathit{Args}$ any variable whose node is labeled with constant $e$.
		}
		\If{Op is \textit{LowByte} and len(Args) = 1 and Args[0] is labeled with a constant below $2^8$}{ 
			\Return{Args[0]}
		}
		\tcc{Many other algebraic rewrite rules}
		\If{Op is commutative}{ 
			Sort $\mathit{Args}$ by textual variable name.
		}
		\For{node $n$ in the E-graph}{ 
			\If{n is labeled with Op and has Args as its edge list}{
				\Return{n}
			}
		}
		$n \gets \textrm{new E-graph node}$\;
		$n.\mathsf{label} \gets \mathit{Op}$\;
		$n.\mathsf{edges} \gets \mathit{Args}$\;
		\Return{$n$}
	}
	\caption{Internalize expression into the E-graph\label{alg:internalize}}
\end{algorithm}
\end{minipage}
    \vspace{-1em}
\end{wrapfigure}

\cref{alg:internalize} presents the \texttt{internalize} algorithm more generally.
Without loss of generality, we assume it is called on an expression that is an operator applied to a list of variables, which are used as E-graph node names.
To internalize expressions with constants and/or deeper nesting of operands, we can simply traverse their trees bottom-up, internalizing each node to obtain a variable (i.e.\ E-graph node) to replace it with.
Elided from the figure are additional rewrite rules to complement the one we include, which notices that an operation to extract the low byte of a constant is extraneous, when the constant is low enough.
The algorithm ends in a linear scan of the E-graph for existing nodes matching the normalized input, which perhaps surprisingly turned out to be more than fast enough for our examples.

The actual implementation includes a general range analysis to establish upper bounds on variable nodes based on bounds on their inputs.
This feature is used to elide truncations whenever appropriate and to gate other rewrite rules: for example, the carry bit resulting from adding small numbers is always 0, and a sum of a couple of carry bits fits in a byte.

Here it is interesting to note which standard features of SMT-based verification tools did \emph{not} need to be implemented, saving us from needing to prove their soundness.
We incorporated no SAT-style management of case splits, nor did we need to include specialized cooperating decision procedures for domains like arithmetic on mathematical integers or bitvectors.
It sufficed to stick to congruence-closure-style reasoning in the theory of equality with uninterpreted functions, augmented with a modest pool of rewrite rules, as SMT solvers are also often effective at applying.
Note also that we omitted a central complexity of E-graph implementations: merging nodes (usually through union-find algorithms) as it is discovered that they are equal.
Our domain is simple enough that relevant equalities are always discovered at node-creation time.
As for symbolic execution, we avoided the characteristic complexities of control flow (e.g.\ merging logical states) and memory access (e.g.\ flexible-enough addressing that pointer aliasing is nontrivial to check).

\subsection{Proof}

We proved the Coq implementation of equivalence checking correct, in the sense that symbolically executing an \assembly (or IR) program produces an expression (E-graph) that would evaluate to the same output as the original program from all starting states.
A fairly direct corollary is our main theorem: if two programs are symbolically executed with the same (symbolic) inputs and produce outputs that are represented by the same E-graph nodes, then the programs are equivalent.
We combine the \fiatir symbolic evaluator with the larger \fiat pipeline, and we verify it against the existing semantics of the IR, before extracting the pipeline to a command-line program.
That program takes as input a choice of cryptographic algorithm, its numeric parameters, and an \assembly file,
and it checks that the \assembly file matches the behavior of the algorithm, by comparing it with the \fiatir code
(itself generated by a verified compiler).

It is important that the top-level theorem of the equivalence checker avoids mentioning any specifics of symbolic execution or E-graphs, as those are relatively complex techniques.
Instead, that theorem refers only to formal semantics of \fiatir and \xassembly.
Slightly more precisely, we depend on the following preconditions, several of which we formalize using notations borrowed from separation logic~\cite{SeplogLICS02}.
\begin{itemize}
\item Calling-convention-designated registers hold input values, input-array base pointers, and output-array base pointers.
\item Input-array base pointers point to arrays in memory holding input values.
\item Output-array base pointers are valid.
\item \texttt{rsp} points to the end of the stack, which must be valid.
\end{itemize}
Then the main theorem concludes the following postconditions.
\begin{itemize}
\item Input-array base pointers are still valid.
\item Output-array base pointers point to arrays in memory holding the output values.
\item The stack base-pointer address is still a valid pointer to an array of the right size.
\item Callee-save registers have the same values as before the code execution.
\item All other memory is untouched.
\end{itemize}

A few other concerns must be stated in the full theorem, including that all source-program initial variable values fit in machine words, arrays are laid out contiguously (the symbolic-execution engine allows more flexible specification of memory contexts), and every array has a start address stored directly in a register.

\section{Performance Evaluation}
\label{s:evaluation}
In this section, we evaluate the performance of code produced by \cryptopt. 
Specifically, we  answer four main questions: 
\begin{enumerate}[nosep]
    \item How does optimization progress over time (\cref{s:evalTime})?
    \item How does \cryptopt compare with traditional compilation (\cref{s:evalTrad})?
    \item Is \cryptopt optimization platform-specific (\cref{s:evalPlatform})?
    \item How does \cryptopt-optimized code perform as part of a cryptographic implementation (\cref{s:evalScmul})?
\end{enumerate}

We first describe the setup and the procedures we use for the evaluation.
We then describe the experiments we carry out and their results.

\clearpage
\subsection{Experimental Setup}\label{s:setup}

In this subsection, we describe the hardware platforms and discuss the 
\emph{generation} of results through optimization, plus the 
\emph{evaluation} of which one is the best of all the results.

\begin{wraptable}{r}{.45\textwidth}
\vspace{-1.5em}
    \scriptsize
    \begin{center}
        \caption{Overview of target machines used in the experiments}
\vspace{-1em}
        \label{tab:machines}

        \begin{tabular}{lll}
            \toprule
            Name & CPU & $\mu$-architecture\\
            \midrule
            1900X  & AMD Ryzen Theadr.~1900X     & Zen 1         \\
            5800X  & AMD Ryzen~7 5800X           & Zen 3         \\
            5950X  & AMD Ryzen~9 5950X           & Zen 3         \\
            7950X  & AMD Ryzen~9 7950X           & Zen 4         \\
            i7 6G  & Intel Core i7-6770HQ        & Skylake-H     \\
            i7 10G & Intel Core i7-10710U        & Comet Lake-U  \\
            i9 10G & Intel Core i9-10900K        & Comet Lake-S  \\
            i7 11G & Intel Core i7-11700KF       & Rocket Lake-S \\
            i9 12G & Intel Core i9-12900KF       & Alder Lake-S  \\
            i9 13G & Intel Core i9-13900KF       & Raptor Lake-S  \\
            \bottomrule
        \end{tabular}
    \end{center}
\vspace{-2em}
\end{wraptable}

\parhead{Hardware Platforms.}
To compare across multiple processor architectures, we evaluate \cryptopt on multiple hardware platforms, summarized in \cref{tab:machines}.
We did not observe differences in optimization behavior on machines with SMT enabled or disabled.
We use Ubuntu Server 22.04.1 LTS for all machines, with all packages up-to-date.
Moreover, because performance counters are not available on the efficiency cores of the i9~12G and i9~13G, we only use the performance cores on those machines.

\begin{wraptable}{r}{.32\textwidth}
\vspace{-1.4em}
    \footnotesize
    \centering
        \caption{Instruction count (average over all eight platforms)}
            \label{tab:avgInstrCount}
            \vspace{-1mm}
        \begin{tabular}{l *{4}{d{3.3}}}
            \toprule
            Primitive & \mc{Multiply} & \mc{Square} \\
            \midrule 
            Curve25519       &  173.300 &  123.333 \\
            NIST P-224       &  226.967 &  222.233 \\
            NIST P-256       &  206.133 &  198.600 \\
            NIST P-384       &  580.400 &  570.300 \\
            SIKEp434         &  968.333 &  927.833 \\
            Curve448         &  550.133 &  359.467 \\
            NIST P-521       &  575.233 &  359.233 \\
            Poly1305         &   73.333 &   55.633 \\
            secp256k1        &  228.333 &  223.500 \\
            \bottomrule
        \end{tabular}
    \vspace{-1em}
\end{wraptable}

\parhead{Generation.}
Every platform has at least four physical cores.
For a fair comparison, we run the same number of optimization processes in parallel on all platforms.
We pin the optimization processes to cores to reduce noise due to context switching.
We choose to run \parallelruns optimizations in parallel, keeping one core free for general OS activity. 
This results in \parallelruns \assembly files for each primitive per platform, of which we report only the best-performing.
The number of assembly instructions varies from 51 to 1281, depending on the field, the operation, and how well the optimization went.
See \cref{tab:avgInstrCount} for details.

\parhead{\barheur.} 
Each optimization process has a budget of \budgett mutations.
In the bet stage, we explore \budgetk initial candidate solutions, optimizing each for \budgeteach mutations.
Hence, overall, we use \budgettone mutations, which are \budgetpercent of the total budget, for the bet part.
The remaining \budgetttwo mutations are used for the run stage of the \barheur strategy.
With those parameters, the generation and verification of all of the 18 \fiatir primitives takes between 20 and 40 wall-clock hours, depending on the machine.

\begin{wraptable}{l}{.33\textwidth}
\vspace{-1em}
    \begin{center}\footnotesize
        \caption{Verification times (average over all ten platforms)}
            \label{tab:verificationTimes}
        \begin{tabular}{l *{4}{d{3.3}}}
            \toprule
            Primitive & \mc{Multiply} & \mc{Square} \\
            \midrule 
            Curve25519       &   0.29~s        &   0.18~s \\
            NIST P-224       &   2.16~s        &   1.94~s \\
            NIST P-256       &   1.07~s        &   0.95~s \\
            NIST P-384       &  29.09~s        &  25.81~s \\
            SIKEp434         & 315.23~s        & 284.32~s \\
            Curve448         &   3.02~s        &   1.44~s \\
            NIST P-521       &   3.14~s        &   1.46~s \\
            Poly1305         &   0.07~s        &   0.06~s \\
            secp256k1        &   1.78~s        &   1.65~s \\
            \bottomrule
        \end{tabular}
    \end{center}
\vspace{-2em}
\end{wraptable}

\parhead{Code Verification.} 
When combined with \fiat, the optimization process verifies that the code it produces is equivalent to the \fiatir code.
We perform verification only on the final output of the optimization. \cref{tab:verificationTimes} shows the time it takes to verify one assembly implementation as an average
over the ten platforms.

\parhead{Performance Metric.}
To compare the performance of different implementations, we need a stable metric.
To reduce system noise we fix the CPU frequency, disable boosting, and set the governor to \texttt{performance}.
We note that we only apply these settings when evaluating the performance but not during optimization.
For more details on this aspect of our experimental setup, see \full{\cref{app:costfunctionevaluation}}{the full version of this paper}.

\subsection{Optimization Progress}
\label{s:evalTime}

The first evaluation question we answer is how the optimization progresses over time.
We include details in \full{\cref{app:evalTime}}{the full version of this paper}, but a summary is that optimization progress is roughly logarithmic in the number of mutations, with some interesting differences in measurement stability across platforms.

\subsection{\cryptopt vs.\ Off-the-Shelf Compilers}
\label{s:evalTrad}

To compare \cryptopt with traditional compilers, we use the implementations of finite-field arithmetic as produced by \fiat.
Specifically, we use \fiat to produce implementations of the multiply and the square functions in nine fields.
We consider prime fields of the standardized NIST P curves:
Curve P-224,
Curve P-256,
Curve P-384 and
Curve P-521~\cite{fips1862}.
Moreover,
we consider the field of the popular `Bitcoin' curve
secp256k1~\cite{sec2},
the high-speed de-facto standard Curve25519~\cite{curve25519},
and a high-security Curve448~\cite{curve448}.
In addition to elliptic-curve cryptography,
we apply our method to the underlying fields of the post-quantum scheme SIKEp434~\cite{sike}
and of the Poly1305 message-authentication scheme~\cite{poly1305}.
It is worth noting that \citet{ErbsenPGSC19} reported that their generated C code was roughly the best-performing available for all elliptic curves, up to the usual vagaries of C-compiler optimizers fluctuating in behavior across versions, so it makes sense to use that C code as our performance baseline.

\begin{wraptable}{r}{0.45\textwidth}
	\small
\vspace{-1em}
	\begin{center}
	\caption{Geometric means of \cryptopt vs.\ off-the-shelf compilers.}
	\label{tab:res-avg}
		\begin{tabular}{@{}lccccc}
		\toprule
			      & \multicolumn{2}{c}{Multiply} & & \multicolumn{2}{c}{Square}\\
			                \cmidrule{2-3}                  \cmidrule{5-6} 
			Curve & Clang & GCC & & Clang & GCC \\
		\midrule
			Curve25519  & \cellcolor{     blue!100}\color{white}{$1.25$} & \cellcolor{     blue!100}\color{white}{$1.16$} &  & \cellcolor{     blue!100}\color{white}{$1.18$} & \cellcolor{     blue!100}\color{white}{$1.17$}\\
			P-224  & \cellcolor{     blue!100}\color{white}{$1.54$} & \cellcolor{     blue!100}\color{white}{$2.52$} &  & \cellcolor{     blue!100}\color{white}{$1.40$} & \cellcolor{     blue!100}\color{white}{$2.56$}\\
			P-256  & \cellcolor{     blue!100}\color{white}{$1.70$} & \cellcolor{     blue!100}\color{white}{$2.61$} &  & \cellcolor{     blue!100}\color{white}{$1.63$} & \cellcolor{     blue!100}\color{white}{$2.59$}\\
			P-384  & \cellcolor{     blue!100}\color{white}{$1.45$} & \cellcolor{     blue!100}\color{white}{$2.49$} &  & \cellcolor{     blue!100}\color{white}{$1.37$} & \cellcolor{     blue!100}\color{white}{$2.51$}\\
			SIKEp434  & \cellcolor{     blue!100}\color{white}{$1.70$} & \cellcolor{     blue!100}\color{white}{$2.43$} &  & \cellcolor{     blue!100}\color{white}{$1.73$} & \cellcolor{     blue!100}\color{white}{$2.39$}\\
			Curve448  & \cellcolor{     blue!100}\color{white}{$1.19$} & \cellcolor{   orange!21 }\color{black}{$0.98$} &  & \cellcolor{     blue!59 }\color{white}{$1.07$} & \cellcolor{     blue!42 }\color{black}{$1.05$}\\
			P-521  & \cellcolor{     blue!100}\color{white}{$1.30$} & \cellcolor{   orange!28 }\color{black}{$0.97$} &  & \cellcolor{     blue!100}\color{white}{$1.35$} & \cellcolor{     blue!27 }\color{black}{$1.03$}\\
			Poly1305  & \cellcolor{     blue!100}\color{white}{$1.12$} & \cellcolor{     blue!100}\color{white}{$1.22$} &  & \cellcolor{     blue!100}\color{white}{$1.11$} & \cellcolor{     blue!100}\color{white}{$1.26$}\\
			secp256k1  & \cellcolor{     blue!100}\color{white}{$1.80$} & \cellcolor{     blue!100}\color{white}{$2.62$} &  & \cellcolor{     blue!100}\color{white}{$1.71$} & \cellcolor{     blue!100}\color{white}{$2.54$}\\
		\bottomrule
		\end{tabular}
	\end{center}
\vspace{-1em}
\end{wraptable}

We run \cryptopt on each of the ten platforms summarized in \cref{tab:machines} and select the best result as described in \cref{s:setup}.
Additionally, we compile the equivalent C code, as produced by \fiat, with GCC~\gccVersion~\cite{gcc} and Clang~\clangVersion~\cite{clang}.
We use the highest optimization level the compilers support and enable native support using the compilation switches \texttt{-march=native -mtune=native -O3}.

\cref{tab:res-avg} shows a summary of the results. 
For each function the table presents the geometric mean performance gain of \cryptopt over GCC and Clang.
The mean is calculated over the different platforms;
see \full{\cref{f:results} in \cref{app:evalPlatform}}{the full version of this paper} for the full details.

The table shows that \cryptopt achieves significant performance gains in the majority of functions.
The performance gains are somewhat more modest when the produced code does not require any memory spills, as is the case for operations in the fields of Curve25519 and Poly1305.
For the large fields, as in P-521 and Curve448, \cryptopt is less successful, achieving modest gains for the square function compared to GCC and slightly underperforming for
multiply compared to GCC.
We note that the code produced for these curves is quite large.
As an example, the resulting \xassembly files for Curve448-mul are in the range of 511--602 instructions and for P-521-mul 509--648 instructions depending on the platform.
For comparison, Curve25519-mul is in the order of 160--195 instructions and in the range of 66--83 for Poly1305-mul.
We suspect that \cryptopt's simple mutations have less impact on the execution time for those big functions than what would be needed to be measurable and direct the optimizer
towards the optimal instruction sequences.
We leave more sophisticated genetic-improvement strategies for future work.

We further evaluated the impact of profile-guided optimization (PGO) on the performance of code produced by the mainstream compilers. 
Specifically, we compiled the methods adding the profiling option ( \texttt{-fprofile-generate} in GCC and \texttt{-fprofile-instr-generate} in Clang). We then ran each function in a tight loop for 10\,000 iterations with random input, generating profile traces.
Lastly, to use the profile traces, we added the \texttt{-fprofile-use} switch (including one call to \texttt{llvm-profdata merge} to create the correct format and the
switch \texttt{-fprofile-instr-use} when using Clang) to our compilation options.
We then measured the performance of the code generated from this last compilation.

Using PGO with Clang only changes the position of the code for the Curve25519 and P-521 fields and only results in negligible performance changes.
Using PGO with GCC improves the mean performance by $\sim$2\% over the whole set of functions.
Notably, PGO improves the performance of the SIKEp434 operations by $\sim$15\%.
In all tested functions, \cryptopt significantly outperforms GCC even when used with PGO.

\subsection{\cryptopt vs.\ Superoptimization}\label{s:cvss}
To compare \cryptopt to superoptimizers, we use STOKE~\cite{Schkufza0A13}.
STOKE supports two operation modes: synthesize, where it aims to generate new code that performs the function; and optimize, where it attempts to modify a function to find a faster alternative.
Synthesize mode failed to generate correct code from scratch even though we let it run for more than three days.
This is expected because in synthesize mode, STOKE aims for small kernels, whereas functions for finite-field arithmetic are on the order of hundreds of instructions.

For the optimize mode,  we compile our test functions with Clang 3.4 and GCC 4.9. (STOKE does not support  newer versions of these compilers.)
We then try to optimize the assembly code that the compilers emit.

In four of the prime fields (Curve25519, Poly1305, P-448 and P-521), the code produced by the compilers uses the \texttt{shrd} instruction, which is not supported in STOKE due to the potential for undefined behavior.
For the remaining fields, in most optimization attempts, STOKE either times out or emits assembly code that contains either syntactical errors that prevent it from being assembled or logical errors that result in incorrect code.
 
We only managed to get results for the square function of secp256k1, when compiled with GCC~4.9.
When optimizing this function, STOKE produces code that is about 6\% faster than the output of compilation with GCC~11. 
However, the code is still 73\% slower than the code that \cryptopt produces for the same function.

\subsection{Platform-Specific Optimization}
\label{s:evalPlatform}

\cryptopt optimizes the execution time of the function on the platform it executes on.
Because different platforms have different hardware components, the fastest code on one platform is not necessarily the fastest on another.
Surprisingly, sometimes it does pay off to optimize on a different platform than the target, apparently when the host supports more stable performance measurement.
See \full{\cref{app:evalPlatform}}{the full version of this paper} for details.

\subsection{Scalar Multiplication}
\label{s:evalScmul}
So far we have focused on the optimized functions in isolation. 
In this section, we investigate the use of the field operations within the context of elliptic-curve cryptography.
Specifically, we investigate implementations of two popular elliptic curves: Curve25519
and secp256k1.
We compare the performance of 15 implementations of these curves, four of which use \cryptopt code for field operations.
In \cref{t:performance}, we summarize the implementations we investigate.

\parhead{State of the Art.}
For Curve25519, the SUPERCOP benchmark framework~\cite{supercop22} provides us with many implementations:
sandy2x~\cite{sandy2x},
amd64-51 and amd64-64~\cite{Chen14},
as well as donna and donna-c64~\cite{curve25519-donna}.
OpenSSL~\cite{openssl} provides three implementations:
a portable C implementation that we identify as O'SSL, and two assembly-based implementations, based on amd64-51 and amd64-64, which we identify as O'SSL fe-51 and O'SSL fe-64, respectively.
At runtime, OpenSSL chooses  which implementation to use, opting by default for O'SSL fe-64.
Additionally, Project Everest \cite{hacl} provides an assembly implementation in which the computations of two field operations are interleaved to achieve better utilization of the CPU pipeline.

For secp256k1, we use
two implementations from the libsecp256k1 library~\cite{libsecp256k1}, one hand-optimized assembly and the other portable C.

\begin{wraptable}{r}{.45\linewidth}
    \vspace{-1.5em}
\footnotesize
\begin{center}
\caption{Performance of Scalar Multiplication (Geometric Mean).\label{t:performance}}

\begin{tabular}{llr}
\toprule
\multicolumn{3}{c}{\textbf{Curve25519}}\\
Implementation                     & Lang  & Cycles      \\
\midrule
sandy2x~\cite{sandy2x}             & asm-v & 486k \\
amd64-64~\cite{Chen14}             & asm   & 542k \\
amd64-51~\cite{Chen14}             & asm   & 546k \\
donna~\cite{curve25519-donna}      & asm-v & 972k \\
donna-c64~\cite{curve25519-donna}  & C     & 577k \\
O'SSL~\cite{openssl}               & C     & 530k \\
O'SSL fe-51~\cite{openssl}         & asm   & 530k \\
O'SSL fe-51+\textbf\cryptopt       & asm   & 524k \\
O'SSL fe-64~\cite{openssl}         & asm   & 455k \\
O'SSL fe-64+\textbf\cryptopt       & asm   & 461k \\
HACL*~fe-64~\cite{hacl}            & asm   & 452k \\
\midrule
 \multicolumn{3}{c}{\textbf{secp256k1}} \\
 Implementation & Lang & Cycles \\
\midrule
libsecp256k1~\cite{libsecp256k1}       & asm & 547k \\
libsecp256k1~\cite{libsecp256k1}       & C   & 530k \\
libsecp256k1+\textbf\cryptopt\ (Fiat)   & asm & 527k \\
libsecp256k1+\textbf\cryptopt\ (Opt)   & asm & 528k \\
\bottomrule
\end{tabular}
\end{center}
    \vspace{-2em}
\end{wraptable}

\parhead{\cryptopt-Based Implementations.}
For the comparison, we use four implementations with \cryptopt-optimized code. 
Specifically, for Curve25519, we replace the field operations in O'SSL-fe51 and in O'SSL-fe64 with \cryptopt-optimized field operations.
We call these implementations O'SSL fe-51+\cryptopt and O'SSL fe-64+\cryptopt, respectively.

For secp256k1 we use two implementations. 
The first, libsecp256k1+\cryptopt (Fiat), uses the scalar multiplication code from \texttt{libsecp256k1} with the field operations as produced by \fiat and optimized with \cryptopt.

Additionally, to demonstrate the use of the \cryptopt optimizer as a stand-alone tool, we use Clang to compile the field operations of the portable implementation of \texttt{libsecp256k1} into LLVM IR, which we convert to the input format of the \cryptopt optimizer. 
We then use the latter to optimize the code, replacing the field operations with the optimized code.
This implementation is called \mbox{libsecp256k1}+\cryptopt (Opt).

\parhead{Evaluation.}
We use the SUPERCOP benchmark framework~\cite{supercop22} to measure the performance of the evaluated implementations.
For each implementation, SUPERCOP tries multiple combinations of compilers and compiler options and reports the execution time (for two base-point multiplications and two
variable-point multiplications) of the fastest compiler setting.
We evaluate each implementation on the ten hardware platforms (c.f.\ \cref{tab:machines}) and report the geometric mean (rounded to the nearest 1000 cycles) in \cref{t:performance}.
(See \full{\cref{f:fulltable2} in \cref{app:detailed}}{the full version of this paper} for the full details.)

\parhead{Results.}
Comparing \cryptopt with the similarly structured hand-optimized implementations of OpenSSL fe-51 and fe-64, we find that the performance with and without \cryptopt is similar.
On average, \cryptopt generates slightly faster implementations for fe-51 and slightly slower implementations for fe-64.
Manual optimization of code requires significant expertise and a large time investment, which needs to be repeated for each finite field.
In contrast, using \cryptopt is fairly straightforward and only requires moderate computing resources to achieve similar results.
\cryptopt also underperforms highly optimized implementations that use a different API (HACL*), which we do not support yet. 

For secp256k1, the libsecp256k1+\cryptopt (Fiat) implementation beats the performance of the hand-tuned assembly, slightly outperforms the C compiled code, and provides verified
formal correctness.
The libsecp256k1+\cryptopt (Opt) implementation achieves higher performance than both the state-of-the-art and our verified implementation albeit slightly slower than the version
based on \fiat.

We leave the tasks of verifying the implementation for secp256k1 and emitting field operations similar to the HACL* API to future work.

\parhead{\cryptopt on New Hardware.}
When looking at the results on specific machines (\full{\cref{f:fulltable2}}{more details in the full version of this paper}), we see that \cryptopt excels on the i7~11G, i9~12G, and i9~13G platforms,
providing the overall fastest implementations for fe-64-based field operations.
On these platforms, \cryptopt-based implementations of Curve25519 and secp256k1 are the fastest, outperforming hand-optimized implementations, including those that use advanced processor features, such as vector instructions. 
The $12^{th}$ generation of Intel processors is a major update of the microarchitecture.
We believe that \cryptopt's automated search allows it to exploit the benefits of the new design automatically.
In contrast, prior implementations and mainstream compilers need to change to adapt to these new features.
We anticipate that in due course, implementations will be adapted to the new design, and hand-tuned implementations will outperform \cryptopt. 
However, \cryptopt does not require manual effort to adapt to new designs.

\subsection{Artifacts}
The artifact, on which the evaluation was done, is available at: \url{https://zenodo.org/record/7710435}, with the DOI \texttt{10.5281/zenodo.7710435}.
The artifact includes instructions to reproduce the claimed results in this paper.
As of April 2023, the most up-to-date version of \fiat can be found in their repository at \url{https://github.com/mit-plv/fiat-crypto}, and up-to-date versions of \cryptopt at
\url{https://github.com/0xADE1A1DE/CryptOpt}.

\section{Conclusion}

We presented \cryptopt, a tool that brings a perhaps-surprising confluence of improving performance and increasing formal assurance.
It tackles the distinctive simplifications and complexities of straight-line cryptographic code.
We showed empirically that certain simplifications to established techniques suffice to set new performance records for important routines on some relevant platforms.
In \emph{generation} of fast code, we developed a simple set of transformation operators that make genetic search effective.
In \emph{checking} of fast code with foundational mechanized proofs, we followed SMT solvers and symbolic-execution engines, while avoiding their most complex aspects, like arithmetic decision procedures or nontrivial pointer-aliasing checks.
We hope that these techniques can be generalized to other domains of compilation.

\begin{acks}
    Input from many anonymous reviewers has helped shaping this paper.
  We are grateful to all for their work and valuable comments.
  In particular, we thank our shepherd, Yaniv David, for  the careful reading,  guidance, and  support.

  This research was supported by 
  the \grantsponsor{GS00}{Air Force Office of Scientific Research (AFOSR)}{} under award number~\grantnum{GS0}{FA9550-20-1-0425};
  the \grantsponsor{GS01}{Australian Research Council}{}  projects~\grantnum{GS01}{DE200101577},~\grantnum{GS01}{DP200102364} and~\grantnum{GS01}{DP210102670};
  the \grantsponsor{GS02}{Blavatnik ICRC at Tel-Aviv University}{};
  \grantsponsor{GS03}{CSIRO's Data61}{};
  the \grantsponsor{GS04}{Deutsche Forschungsgemeinschaft (DFG, Germany's Excellence Strategy) under Germany's Excellence Strategy}{}~\grantnum{GS04}{EXC 2092 CASA - 390781972};
  the \grantsponsor{GS05}{National Science Foundation}{} under grants~\grantnum{GS05}{CNS-1954712 and CNS-2130671};
  the \grantsponsor{GS06}{National Science Foundation Expedition on the Science of Deep Specification}{} (award~\grantnum{GS06}{CCF-1521584});
  the \grantsponsor{GS07}{Phoenix HPC service at the University of Adelaide}{};
  and gifts from
  \grantsponsor{GS08}{Amazon Web Services}{},
  \grantsponsor{GS09}{AMD}{},
  \grantsponsor{GS10}{Facebook}{},
  \grantsponsor{GS11}{Google}{},
  \grantsponsor{GS12}{Intel}{} and the
  \grantsponsor{GS13}{Tezos Foundation}{}.

Part of this work was carried out while Chitchanok Chuengsatiansup, Markus Wagner, and Yuval Yarom were affiliated with the University of Adelaide, and while Chuyue Sun was
affiliated with the Massachusetts Institute of Technology.

\end{acks}

\clearpage
\bibliography{cryptopt}


\ifFull
\appendix

\section{On Reliable Performance Measurement}
\label{app:costfunctionevaluation}

\subsection{Background}

The execution time of a program can be affected by a myriad of factors, such as,
the state of cache memory~\cite{variabilityCPU};
context switches~\cite{threadContextSwitch};
memory layout and OS environment variables~\cite{stabilizer,producingWrongData}.
Even when a platform is rebooted between runs there can be significant variation in benchmark run times~\cite{InitialSystemState}.
Furthermore, measurements of the benchmark operation can be subject to inherent and external random changes to the benchmark operation,
as well as the system state prior to execution~\cite{rigorousSystemStates,benchmarkPrecision}.
In addition, sensor readings themselves can drift over time~\cite{bokhari2019mind}.
Consequently, all these factors together result in measurement noise, which can affect both the sensitivity and specificity of performance measurements.
One recent work, which surveyed and compared validation approaches in the context of energy consumption optimization~\cite{Bokhari2020},
found that a number of measurement approaches failed to mitigate the effects of measurement noise.
Only the proposed approach \approachRRR was not misled:
it exercised the tests in a rotated-round-robin fashion, while accommodating the necessary regular restarting and recharging of the test platform.

\subsection{Cost-Function Evaluation}

Recall that \cryptopt generates solutions, mutates them, and measures their respective performance.
Measuring this performance on physical machines is inherently noisy.
To lower the effect of noise and enable more stable and fair comparisons,
we base our measurement on \approachRRR, which exercises tests in a rotated-round-robin fashion.
This approach was used in previous work~\cite{Bokhari2020} in the context of energy consumption in order to mitigate the effects of measurement noise. 

In our work, we adapt \approachRRR in two ways.
First, we forgo the restart of the computer as we do not observe any measurement drift over time.
Second, we perform a random scheduling of program variants for performance measurement instead of strictly following a given order of measurements.

\begin{algorithm}[t]
    \SetKwInOut{Input}{input}
    \SetKwInOut{Output}{output}
    \SetKwProg{Function}{function}{ \\begin}{end}
    \SetKwFor{RepTimes}{repeat}{times}{end}
    \SetKwBlock{Block}{\{}{\}}
    \Input{$\mathit{A}$ pointer to code A,\newline $\mathit{B}$ pointer to code B,\newline $\mathit{bs}$ size of a measurement batch,\newline $\mathit{nob}$ number of batches to measure}
    \Output{$\mathit{PA}$ performance of code A in cycles,\newline $\mathit{PB}$ performance of code B in cycles}
    \BlankLine
    \Function{measure($\mathit{A}$, $\mathit{B}$, $\mathit{bs}$, $\mathit{nob}$) }{

        \tcc{initialize cycle lists}
        $cycles_A \gets \textrm{empty list}$\\
        $cycles_B \gets \textrm{empty list}$\\

        \Repeat{both have been measured nob times}{

            \tcc{f points to either A or B}
            $f \gets \textrm{randomSelect}(A,\ B)$\\

            \BlankLine
            \tcc{run bs times and get elapsed cycles}
            $\mathit{cycles} \gets \textrm{countCyclesForNRuns}(bs, f)$\\

            \BlankLine
            \tcc{append elapsed cycles to list A/B resp.}
            $\textrm{append}(\mathit{cycles}_f, \mathit{cycles})$\\
        }

        \BlankLine
        $\mathit{PA} \gets \textrm{median}(\mathit{cycles}_A)$\\
        $\mathit{PB} \gets \textrm{median}(\mathit{cycles}_B)$\\
        \BlankLine
        \Return{$\mathit{PA}, \mathit{PB}$}
    }
    \caption{Execution Time Measurement}
    \label{alg:r3}
\end{algorithm}

\parhead{Measuring Performance.}
\cref{alg:r3} presents our performance-measurement routine.
We now explain the rationale behind this from the inside out (inside the loop of the algorithm to the outside of the measure function).

Cycle counters are integers with a granularity of at best $1$.
Recall that \cryptopt can optimize functions of arbitrary size, ranging in our experiments from fewer than 100 to around 2000 instructions.
That means that the accuracy of the fixed-granularity counter relative to the length-varying function changes.
In other words, the same ``stopwatch'' is not appropriate to time an Olympic sprinter and a marathon runner.

The solution to that is to measure multiple executions of the same function instead of just one.
We call multiple executions of the same function a \emph{batch}.
The batch size $\mathit{bs}$ specifies how many executions of the same function are being measured with one measurement.
It follows logically that small functions should have a bigger $\mathit{bs}$ than long functions in order to get similar cycle accuracy.

Hence, we came up with the idea of the $\mathit{cyclegoal}$.
That is, we want the measured cycles count for one batch across all functions to be in the order of $\mathit{cyclegoal}=10~000$ measured cycles.
The reason for this particular number is that we then have four to five digits of accuracy.
\cryptopt then scales the value for $\mathit{bs}$ up or down by comparing the measured cycles against the $\mathit{cyclegoal}$.

This technique also mitigates another inherent problem that arises when measuring performance in cycles on hardware: 
different hardware platforms interpret ``cycles'' differently, and additional challenges arise due to different types of boosting.
However, because \cryptopt adjusts $\mathit{bs}$ in each evaluation of the mutation (i.e.\ before each call to ``measure'') and it does so on each platform independently,
we get comparable accuracy for all functions across platforms.

Inherent with measuring time on hardware is measurement noise, which we as developers can hardly control.
This noise can be due to the OS's interrupt event handlers or process scheduling.
It can also be due to temperature rises the hardware limits itself.
The technique above mitigates the problems having a bad ``stopwatch'' but at the same time increases the captured noise.
Similar to the \approachRRR, we mitigate this problem by simply measuring each batch multiple times, called \emph{numbers of batches} or $\mathit{nob}$ for short.
That means that we have $\mathit{nob}$ measurements for the same function.
We take the median as the performance measurement for a function to drop outlier measurements.
(Using the minimum rather than median did not have statistically significant effects -- neither positive nor negative.)
Empirically, setting $\mathit{nob}=31$ works well on our experimental platforms. 

Complex processors nowadays can cause biases to one or another function by their speculative execution, prediction behavior and caching, just to name a few reasons.
The \approachRRR also mitigates this issue by using a random sequence of the functions to measure.
We do this by randomly selecting either function to measure a batch of.

\subsection{\assemblyline}

\cryptopt uses \assemblyline~\cite{assemblyline} to assemble the generated initial and mutated code into memory positions~$\mathit{A}$ and $\mathit{B}$.
\assemblyline is a lightweight in-memory assembler for \xassembly instructions.
It takes an input string of instructions, assembles them into machine code and returns a pointer to the executable code.
We use \assemblyline to eliminate the overhead of writing to the file system and invoking a typical compiler tool chain for every evaluation.
Additionally, \cryptopt uses \assemblyline to assemble the code to the start of a memory page.
This reduces noise in measurements because it always assembles aligned code for all functions,
which, in turn, reduces memory-biased performance impacts~\cite{producingWrongData}.
Once assembled, \cryptopt calls the \textsf{measure} function with pointers to that code and values for batch size $\mathit{bs}$ and number of batches $\mathit{nob}$.

\section{Optimization Progress}
\label{app:evalTime}

\newcommand{\archA}{i7~10G\xspace}
\newcommand{\archB}{1900X\xspace}
\newcommand{\maxgain}{80\xspace}
\newcommand{\mingain}{61\xspace}

\newcommand{\cc}{Clang\xspace}

\begin{figure}[t]
    \begin{subfigure}[t]{\linewidth}
        \input{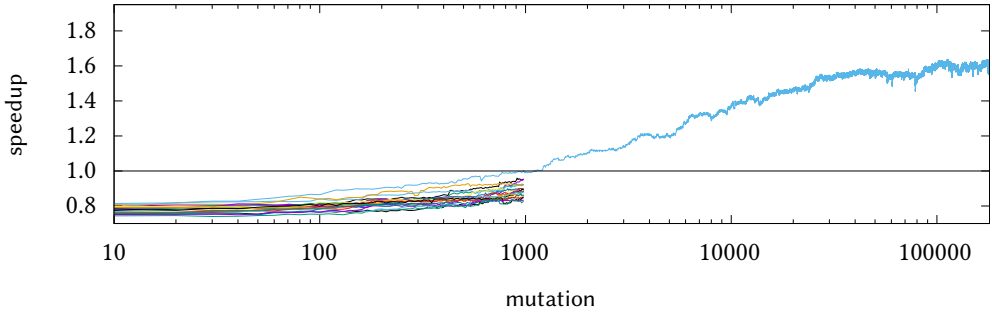}
        \caption{NIST~P-384--square on \archA}
    \end{subfigure}
    \begin{subfigure}[t]{\linewidth}
        \input{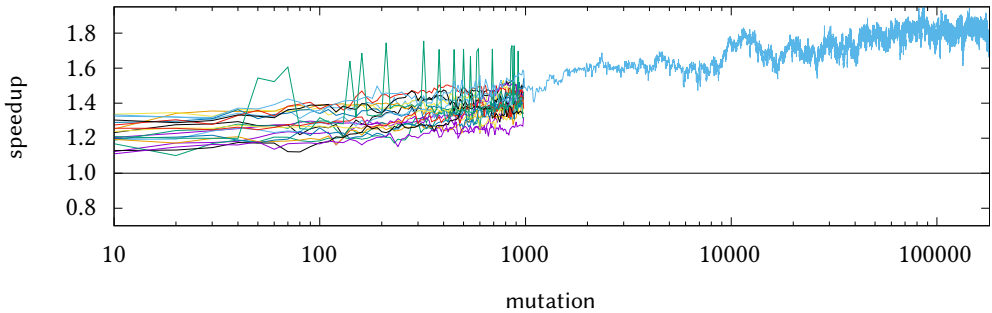}
        \caption{secp256k1--mul on \archB}
    \end{subfigure}
    \caption{\cryptopt optimization progress of NIST~P-384--square and secp256k1--mul on two different platforms. Each line traces an optimization run over time (X-axis, log-scale), showing the relative
        performance gain over \cc. \barheur used for \budgetk runs with a budget of \budgeteach mutations each. The best-performing candidate continues for another \budgetttwo
        mutations. Progress over time is roughly logarithmic in the number of mutations, achieving a performance gain of \mingain--\maxgain{}\%.
        We generate one data point every ten mutations.
    }
    \label{f:conv}
\end{figure}

The first evaluation question we answer is how the optimization progresses over time.
\cref{f:conv} shows examples of optimization runs on \archA and \archB, targeting the field-square operation corresponding to the prime field for NIST~P-384 and the
field-multiplication for the prime field secp256k1.

The figures show performance over time, where performance is measured as gain over the baseline \cc-compiled code, while optimization time is measured in mutations (shown in log scale).

The figure shows the \barheur strategy in action.
The optimization starts with \budgetk runs, each with a budget of \budgeteach mutations.
\cryptopt then selects the best-performing and continues optimizing it for a further \budgetttwo mutations.
As the figure shows, in all runs, optimization progress is roughly logarithmic in the number of mutations. 

A notable difference between the two optimizations is the overall performance gain and the presence of noise.
On the \archA platform the optimization converges rather quickly, whereas on
\archB we see monotonic improvement until the very end.
Finally, we note that these variations and performance gain are dependent on field, method and machine; here we see a gain \mingain--\maxgain{}\%.

\section{Platform-Specific Optimization}
\label{app:evalPlatform}

\renewcommand{\arraystretch}{0.45}
\begin{table*}\centering
\caption{\tiny Optimization results. We show the relative improvements in \% for the multiplication (top) and squaring (bottom) operations; time savings are marked in blue. 
First, to observe hardware-specific optimization, the 10-by-10 matrix shows the performance the optimized operation that have been optimized on one machine and then run on another. 
The subsequent two rows (Clang/GCC) then show the time savings of our optimized operations over off-the-shelf-compilers.
Lastly, ``Final'' shows the time savings of our best-performing implementation over the best-performing compiler-generated version.
}
\label{f:results}

	\setlength{\tabcolsep}{2.7px}
	\begin{tabular}{ @{}r ccccccccccc}
		\addlinespace[-1em]
		\multicolumn{12}{c}{\small Curve25519}\\
		\addlinespace[-.1em]

		\cmidrule{1-12}
		\fontsize{5}{7}\selectfont run~on & \fontsize{5}{7}\selectfont\rotatebox{90}{1900X} & \fontsize{5}{7}\selectfont\rotatebox{90}{5800X} & \fontsize{5}{7}\selectfont\rotatebox{90}{5950X} & \fontsize{5}{7}\selectfont\rotatebox{90}{7950X} & \fontsize{5}{7}\selectfont\rotatebox{90}{i7 6G} & \fontsize{5}{7}\selectfont\rotatebox{90}{i7 10G} & \fontsize{5}{7}\selectfont\rotatebox{90}{i9 10G} & \fontsize{5}{7}\selectfont\rotatebox{90}{i7 11G} & \fontsize{5}{7}\selectfont\rotatebox{90}{i9 12G} & \fontsize{5}{7}\selectfont\rotatebox{90}{i9 13G} & \fontsize{5}{7}\selectfont\rotatebox{90}{G.M.}\\
		\addlinespace[-.1em]

		\fontsize{5}{7}\selectfont opt~on  & & & & & & & & & & & \\
\end{tabular}
	\setlength{\tabcolsep}{2.7px}
	\begin{tabular}{ @{}r ccccccccccc}
		\addlinespace[-1em]
		\multicolumn{12}{c}{\small P-224}\\
		\addlinespace[-.1em]

		\cmidrule{1-12}
		\fontsize{5}{7}\selectfont run~on & \fontsize{5}{7}\selectfont\rotatebox{90}{1900X} & \fontsize{5}{7}\selectfont\rotatebox{90}{5800X} & \fontsize{5}{7}\selectfont\rotatebox{90}{5950X} & \fontsize{5}{7}\selectfont\rotatebox{90}{7950X} & \fontsize{5}{7}\selectfont\rotatebox{90}{i7 6G} & \fontsize{5}{7}\selectfont\rotatebox{90}{i7 10G} & \fontsize{5}{7}\selectfont\rotatebox{90}{i9 10G} & \fontsize{5}{7}\selectfont\rotatebox{90}{i7 11G} & \fontsize{5}{7}\selectfont\rotatebox{90}{i9 12G} & \fontsize{5}{7}\selectfont\rotatebox{90}{i9 13G} & \fontsize{5}{7}\selectfont\rotatebox{90}{G.M.}\\
		\addlinespace[-.1em]

		\fontsize{5}{7}\selectfont opt~on  & & & & & & & & & & & \\
\end{tabular}
	\setlength{\tabcolsep}{2.7px}
	\begin{tabular}{ @{}r ccccccccccc}
		\addlinespace[-1em]
		\multicolumn{12}{c}{\small P-256}\\
		\addlinespace[-.1em]

		\cmidrule{1-12}
		\fontsize{5}{7}\selectfont run~on & \fontsize{5}{7}\selectfont\rotatebox{90}{1900X} & \fontsize{5}{7}\selectfont\rotatebox{90}{5800X} & \fontsize{5}{7}\selectfont\rotatebox{90}{5950X} & \fontsize{5}{7}\selectfont\rotatebox{90}{7950X} & \fontsize{5}{7}\selectfont\rotatebox{90}{i7 6G} & \fontsize{5}{7}\selectfont\rotatebox{90}{i7 10G} & \fontsize{5}{7}\selectfont\rotatebox{90}{i9 10G} & \fontsize{5}{7}\selectfont\rotatebox{90}{i7 11G} & \fontsize{5}{7}\selectfont\rotatebox{90}{i9 12G} & \fontsize{5}{7}\selectfont\rotatebox{90}{i9 13G} & \fontsize{5}{7}\selectfont\rotatebox{90}{G.M.}\\
		\addlinespace[-.1em]

		\fontsize{5}{7}\selectfont opt~on  & & & & & & & & & & & \\
\end{tabular}
	\setlength{\tabcolsep}{2.7px}
	\begin{tabular}{ @{}r ccccccccccc}
		\addlinespace[-1em]
		\multicolumn{12}{c}{\small P-384}\\
		\addlinespace[-.1em]

		\cmidrule{1-12}
		\fontsize{5}{7}\selectfont run~on & \fontsize{5}{7}\selectfont\rotatebox{90}{1900X} & \fontsize{5}{7}\selectfont\rotatebox{90}{5800X} & \fontsize{5}{7}\selectfont\rotatebox{90}{5950X} & \fontsize{5}{7}\selectfont\rotatebox{90}{7950X} & \fontsize{5}{7}\selectfont\rotatebox{90}{i7 6G} & \fontsize{5}{7}\selectfont\rotatebox{90}{i7 10G} & \fontsize{5}{7}\selectfont\rotatebox{90}{i9 10G} & \fontsize{5}{7}\selectfont\rotatebox{90}{i7 11G} & \fontsize{5}{7}\selectfont\rotatebox{90}{i9 12G} & \fontsize{5}{7}\selectfont\rotatebox{90}{i9 13G} & \fontsize{5}{7}\selectfont\rotatebox{90}{G.M.}\\
		\addlinespace[-.1em]

		\fontsize{5}{7}\selectfont opt~on  & & & & & & & & & & & \\
\end{tabular}
	\setlength{\tabcolsep}{2.7px}
	\begin{tabular}{ @{}r ccccccccccc}
		\addlinespace[-1em]
		\multicolumn{12}{c}{\small SIKEp434}\\
		\addlinespace[-.1em]

		\cmidrule{1-12}
		\fontsize{5}{7}\selectfont run~on & \fontsize{5}{7}\selectfont\rotatebox{90}{1900X} & \fontsize{5}{7}\selectfont\rotatebox{90}{5800X} & \fontsize{5}{7}\selectfont\rotatebox{90}{5950X} & \fontsize{5}{7}\selectfont\rotatebox{90}{7950X} & \fontsize{5}{7}\selectfont\rotatebox{90}{i7 6G} & \fontsize{5}{7}\selectfont\rotatebox{90}{i7 10G} & \fontsize{5}{7}\selectfont\rotatebox{90}{i9 10G} & \fontsize{5}{7}\selectfont\rotatebox{90}{i7 11G} & \fontsize{5}{7}\selectfont\rotatebox{90}{i9 12G} & \fontsize{5}{7}\selectfont\rotatebox{90}{i9 13G} & \fontsize{5}{7}\selectfont\rotatebox{90}{G.M.}\\
		\addlinespace[-.1em]

		\fontsize{5}{7}\selectfont opt~on  & & & & & & & & & & & \\
\end{tabular}
	\setlength{\tabcolsep}{2.7px}
	\begin{tabular}{ @{}r ccccccccccc}
		\addlinespace[-1em]
		\multicolumn{12}{c}{\small Curve448}\\
		\addlinespace[-.1em]

		\cmidrule{1-12}
		\fontsize{5}{7}\selectfont run~on & \fontsize{5}{7}\selectfont\rotatebox{90}{1900X} & \fontsize{5}{7}\selectfont\rotatebox{90}{5800X} & \fontsize{5}{7}\selectfont\rotatebox{90}{5950X} & \fontsize{5}{7}\selectfont\rotatebox{90}{7950X} & \fontsize{5}{7}\selectfont\rotatebox{90}{i7 6G} & \fontsize{5}{7}\selectfont\rotatebox{90}{i7 10G} & \fontsize{5}{7}\selectfont\rotatebox{90}{i9 10G} & \fontsize{5}{7}\selectfont\rotatebox{90}{i7 11G} & \fontsize{5}{7}\selectfont\rotatebox{90}{i9 12G} & \fontsize{5}{7}\selectfont\rotatebox{90}{i9 13G} & \fontsize{5}{7}\selectfont\rotatebox{90}{G.M.}\\
		\addlinespace[-.1em]

		\fontsize{5}{7}\selectfont opt~on  & & & & & & & & & & & \\
\end{tabular}
	\setlength{\tabcolsep}{2.7px}
	\begin{tabular}{ @{}r ccccccccccc}
		\addlinespace[-1em]
		\multicolumn{12}{c}{\small P-521}\\
		\addlinespace[-.1em]

		\cmidrule{1-12}
		\fontsize{5}{7}\selectfont run~on & \fontsize{5}{7}\selectfont\rotatebox{90}{1900X} & \fontsize{5}{7}\selectfont\rotatebox{90}{5800X} & \fontsize{5}{7}\selectfont\rotatebox{90}{5950X} & \fontsize{5}{7}\selectfont\rotatebox{90}{7950X} & \fontsize{5}{7}\selectfont\rotatebox{90}{i7 6G} & \fontsize{5}{7}\selectfont\rotatebox{90}{i7 10G} & \fontsize{5}{7}\selectfont\rotatebox{90}{i9 10G} & \fontsize{5}{7}\selectfont\rotatebox{90}{i7 11G} & \fontsize{5}{7}\selectfont\rotatebox{90}{i9 12G} & \fontsize{5}{7}\selectfont\rotatebox{90}{i9 13G} & \fontsize{5}{7}\selectfont\rotatebox{90}{G.M.}\\
		\addlinespace[-.1em]

		\fontsize{5}{7}\selectfont opt~on  & & & & & & & & & & & \\
\end{tabular}
	\setlength{\tabcolsep}{2.7px}
	\begin{tabular}{ @{}r ccccccccccc}
		\addlinespace[-1em]
		\multicolumn{12}{c}{\small Poly1305}\\
		\addlinespace[-.1em]

		\cmidrule{1-12}
		\fontsize{5}{7}\selectfont run~on & \fontsize{5}{7}\selectfont\rotatebox{90}{1900X} & \fontsize{5}{7}\selectfont\rotatebox{90}{5800X} & \fontsize{5}{7}\selectfont\rotatebox{90}{5950X} & \fontsize{5}{7}\selectfont\rotatebox{90}{7950X} & \fontsize{5}{7}\selectfont\rotatebox{90}{i7 6G} & \fontsize{5}{7}\selectfont\rotatebox{90}{i7 10G} & \fontsize{5}{7}\selectfont\rotatebox{90}{i9 10G} & \fontsize{5}{7}\selectfont\rotatebox{90}{i7 11G} & \fontsize{5}{7}\selectfont\rotatebox{90}{i9 12G} & \fontsize{5}{7}\selectfont\rotatebox{90}{i9 13G} & \fontsize{5}{7}\selectfont\rotatebox{90}{G.M.}\\
		\addlinespace[-.1em]

		\fontsize{5}{7}\selectfont opt~on  & & & & & & & & & & & \\
\end{tabular}
	\setlength{\tabcolsep}{2.7px}
	\begin{tabular}{ @{}r ccccccccccc}
		\addlinespace[-1em]
		\multicolumn{12}{c}{\small secp256k1}\\
		\addlinespace[-.1em]

		\cmidrule{1-12}
		\fontsize{5}{7}\selectfont run~on & \fontsize{5}{7}\selectfont\rotatebox{90}{1900X} & \fontsize{5}{7}\selectfont\rotatebox{90}{5800X} & \fontsize{5}{7}\selectfont\rotatebox{90}{5950X} & \fontsize{5}{7}\selectfont\rotatebox{90}{7950X} & \fontsize{5}{7}\selectfont\rotatebox{90}{i7 6G} & \fontsize{5}{7}\selectfont\rotatebox{90}{i7 10G} & \fontsize{5}{7}\selectfont\rotatebox{90}{i9 10G} & \fontsize{5}{7}\selectfont\rotatebox{90}{i7 11G} & \fontsize{5}{7}\selectfont\rotatebox{90}{i9 12G} & \fontsize{5}{7}\selectfont\rotatebox{90}{i9 13G} & \fontsize{5}{7}\selectfont\rotatebox{90}{G.M.}\\
		\addlinespace[-.1em]

		\fontsize{5}{7}\selectfont opt~on  & & & & & & & & & & & \\
\end{tabular}
\end{table*}
\cryptopt optimizes the execution time of the function on the platform it executes on.
Because different platforms have different hardware components, the fastest code on one platform is not necessarily the fastest on another.
Thus, in this section we ask whether \cryptopt overfits to the platform it executes on, as opposed to universally optimizing for all platforms.
That is, we test how \emph{native optimization}, where the code executes on the same platform it was optimized on,
compares with \emph{cross optimization}, where code is optimized on one platform and executes on another.

To test this, we first optimize each of the functions on each of the platforms.
We then measure the performance of the produced code from all platforms on all platforms (as per evaluation described in \cref{app:costfunctionevaluation}).
\cref{f:results} shows the results for the operations multiply and square on all nine fields we tested.
Rows correspond to the platform the code was optimized on and columns to the platform the code is executed on.
The value in a cell shows the ratio between the execution time of the cross-optimized code and that of the native code.
An example value of $1.10$ indicates that the cross-platform code (from the ``row'' platform) needs $1.10$ times as many cycles as the native code (from the ``column'' platform).
In other words, a blue cell means that code from the column architecture outperforms code from the row architecture, and the number indicates by how much.

The additional column \gm shows the geometric mean of the numbers in the row.
This gives a measure of ``how platform-specific'' one optimization is:
A value larger than one (blue) in column \gm in the row for platform X indicates that on average, code from other platforms outperform the code from X, if ran on platforms other than X.
In turn, a value smaller than one means that the code from platform X tends to outperform platform-specific code on other platforms.
Please note that a ``large number'' can arise for two reasons:
(a) The platform-specific code is very bad (thus easy to outperform),
or (b) the generated code is very generic and can thus compete with the platform-specific ones.
To see which is applicable per platform, one needs to see how well the code compares to \ots compilers.
Note that this is also not a universal measure, as \ots compilers generate different code per platform.
Alternatively, running the same code on all platforms would also not give a fair baseline,
as compilers would not be enabled to generate platform-specific code, whereas \cryptopt would be.

As can be seen in \cref{f:results} for Curve25519, optimizing and running on the same architecture tends to outperform cross-architecture optimizations (which
would mean the 10-by-10 is mostly blue, and the \gm column is fully blue, too).
However, there are cases where cross-architectural optimizations are better.
In particular it appears that optimization of SIKEp434 functions on AMD is not ideal, and the results underperform code optimized on Intel processors.
(Still, faster than compiling with \ots compilers).

Another effect that the table highlights is that optimization overfits for processor families, 
for some families more than others.
In particular, Intel 6th and 10th generations show little differences in the respective inter-platform performance.
For example, the related 3x3 sub-matrices for Curve25519 in \cref{f:results} have values in the small interval $[0.98, 1.02]$ (mul), $[0.97, 1.03]$ (square);
as do AMD Zen~3 processors in the 2x2 sub-matrices have values in the small interval $[0.97, 1.05]$ (mul), and $[1.00, 1.02]$ for square.

This specialization to platforms is a double-edged sword. 
On the one hand, it allows \cryptopt to produce high-performing code, as previously shown.
On the other hand, it also means that the code is potentially less generic, and it might be less performant if executed on platforms other than originally intended, as observed in
our a-posteriori analysis.

\section{Detailed Performance Information}\label{app:detailed}
\cref{f:fulltable2} shows the detailed performance information of the scalar multiplication experiments.
We use SUPERCOP to evaluate and report the cycles for two base-point and two variable-point scalar multiplications.
The table is divided in two sections, one for Curve25519 and one for secp256k1.
In each of those sections we compare different implementations of the scalar multiplication for respective curve against each other.
The language, in which the core operations are implemented, is written in the column \emph{Lang}.
The following columns show the elapsed cycles for the scalar multiplication.
The fastest implementation per section and per platform is highlighted in \textbf{(bold text)} and notes a \textbf{(1.00x)}.
All other implementations are then compared against this fastest in the form of the ratio, which is written in parenthesis.
E.g. 1.05x means that it takes 1.05 times as many cycles than the fastest.
The last column, \gm, is the geometric mean of the cycles across all platforms, the ratio is then recalculated on that \gm as well.

For secp256k1, we base all evaluation on the implementation for scalar multiplication on the one from \texttt{libsecp256k1}.
What changes is the implementation of the field arithmetic.
That is, the first two rows are the asm and C implementations from \texttt{libsecp256k1}.
In the third row, we plug in verified and optimized field operations based on \fiat,
and in the last row we use the optimized versions of \texttt{libsecp256k1}'s own implementation (i.e. no formal correctness verification).

Note:
``ots'' stands for off-the-shelf; ``asm'' means assembly; ``-v'' indicates the use of vector instructions.
The HACL* implementation \cite{hacl} uses parallelized field arithmetic.
For Curve25519, ``-51`` and ``-64`` indicates whether the representation for the field elements is unsaturated or saturated.

\afterpage{%
    \clearpage%
    \begin{landscape}
        \tiny
        \centering%
        \newcommand{\rowA}{\cellcolor[rgb]{0.9,0.9,0.9}}
\newcommand{\rowB}{\cellcolor[rgb]{1.0,1.0,1.0}}
\begin{tabular}{cllrrrrrrrrrrr}
\toprule
 & Implementation                                   &                Lang. &                                              1900X &                                              5800X &                                              5950X &                                              7950X &                                              i7 6G &                                             i7 10G &                                             i9 10G &                                             i7 11G &                                             i9 12G &                                             i9 13G &                                         G.M.\\
\midrule
\multirow{11}{*}[-2em]{\rotatebox[origin=c]{90}{\centering Curve25519}}
 & \rowA sandy2x~\cite{sandy2x}                     & \rowA          asm-v & \rowA           603k {\tiny (1.05x)}\hspace{-.5em} & \rowA  \textbf{426k {\tiny (1.00x)}}\hspace{-.5em} & \rowA  \textbf{426k {\tiny (1.00x)}}\hspace{-.5em} & \rowA  \textbf{421k {\tiny (1.00x)}}\hspace{-.5em} & \rowA           521k {\tiny (1.15x)}\hspace{-.5em} & \rowA           521k {\tiny (1.15x)}\hspace{-.5em} & \rowA           521k {\tiny (1.15x)}\hspace{-.5em} & \rowA           492k {\tiny (1.15x)}\hspace{-.5em} & \rowA           480k {\tiny (1.31x)}\hspace{-.5em} & \rowA           480k {\tiny (1.31x)}\hspace{-.5em} & \rowA           486k {\tiny (1.07x)}\hspace{-.5em} \\
 & \rowB amd64-64~\cite{Chen14}                     & \rowB            asm & \rowB           622k {\tiny (1.09x)}\hspace{-.5em} & \rowB           573k {\tiny (1.34x)}\hspace{-.5em} & \rowB           573k {\tiny (1.34x)}\hspace{-.5em} & \rowB           528k {\tiny (1.25x)}\hspace{-.5em} & \rowB           569k {\tiny (1.26x)}\hspace{-.5em} & \rowB           571k {\tiny (1.26x)}\hspace{-.5em} & \rowB           572k {\tiny (1.26x)}\hspace{-.5em} & \rowB           536k {\tiny (1.26x)}\hspace{-.5em} & \rowB           452k {\tiny (1.23x)}\hspace{-.5em} & \rowB           452k {\tiny (1.23x)}\hspace{-.5em} & \rowB           542k {\tiny (1.20x)}\hspace{-.5em} \\
 & \rowA amd64-51~\cite{Chen14}                     & \rowA            asm & \rowA           772k {\tiny (1.35x)}\hspace{-.5em} & \rowA           586k {\tiny (1.37x)}\hspace{-.5em} & \rowA           586k {\tiny (1.37x)}\hspace{-.5em} & \rowA           563k {\tiny (1.34x)}\hspace{-.5em} & \rowA           567k {\tiny (1.26x)}\hspace{-.5em} & \rowA           567k {\tiny (1.26x)}\hspace{-.5em} & \rowA           567k {\tiny (1.25x)}\hspace{-.5em} & \rowA           490k {\tiny (1.15x)}\hspace{-.5em} & \rowA           420k {\tiny (1.14x)}\hspace{-.5em} & \rowA           421k {\tiny (1.15x)}\hspace{-.5em} & \rowA           546k {\tiny (1.21x)}\hspace{-.5em} \\
 & \rowB donna~\cite{curve25519-donna}              & \rowB          asm-v & \rowB          1089k {\tiny (1.90x)}\hspace{-.5em} & \rowB           973k {\tiny (2.28x)}\hspace{-.5em} & \rowB           971k {\tiny (2.28x)}\hspace{-.5em} & \rowB           954k {\tiny (2.26x)}\hspace{-.5em} & \rowB          1023k {\tiny (2.27x)}\hspace{-.5em} & \rowB          1015k {\tiny (2.25x)}\hspace{-.5em} & \rowB          1014k {\tiny (2.24x)}\hspace{-.5em} & \rowB           958k {\tiny (2.25x)}\hspace{-.5em} & \rowB           871k {\tiny (2.37x)}\hspace{-.5em} & \rowB           872k {\tiny (2.38x)}\hspace{-.5em} & \rowB           972k {\tiny (2.15x)}\hspace{-.5em} \\
 & \rowA donna-c64~\cite{curve25519-donna}          & \rowA              C & \rowA           807k {\tiny (1.41x)}\hspace{-.5em} & \rowA           599k {\tiny (1.40x)}\hspace{-.5em} & \rowA           598k {\tiny (1.40x)}\hspace{-.5em} & \rowA           637k {\tiny (1.51x)}\hspace{-.5em} & \rowA           581k {\tiny (1.29x)}\hspace{-.5em} & \rowA           582k {\tiny (1.29x)}\hspace{-.5em} & \rowA           582k {\tiny (1.29x)}\hspace{-.5em} & \rowA           550k {\tiny (1.29x)}\hspace{-.5em} & \rowA           453k {\tiny (1.23x)}\hspace{-.5em} & \rowA           451k {\tiny (1.23x)}\hspace{-.5em} & \rowA           577k {\tiny (1.27x)}\hspace{-.5em} \\
 & \rowB OSSL ots~\cite{openssl}                    & \rowB              C & \rowB           700k {\tiny (1.22x)}\hspace{-.5em} & \rowB           538k {\tiny (1.26x)}\hspace{-.5em} & \rowB           541k {\tiny (1.27x)}\hspace{-.5em} & \rowB           536k {\tiny (1.27x)}\hspace{-.5em} & \rowB           568k {\tiny (1.26x)}\hspace{-.5em} & \rowB           567k {\tiny (1.26x)}\hspace{-.5em} & \rowB           568k {\tiny (1.25x)}\hspace{-.5em} & \rowB           494k {\tiny (1.16x)}\hspace{-.5em} & \rowB           421k {\tiny (1.15x)}\hspace{-.5em} & \rowB           420k {\tiny (1.15x)}\hspace{-.5em} & \rowB           530k {\tiny (1.17x)}\hspace{-.5em} \\
 & \rowA OSSL fe-51 ots~\cite{openssl}              & \rowA            asm & \rowA           696k {\tiny (1.22x)}\hspace{-.5em} & \rowA           540k {\tiny (1.27x)}\hspace{-.5em} & \rowA           540k {\tiny (1.27x)}\hspace{-.5em} & \rowA           536k {\tiny (1.27x)}\hspace{-.5em} & \rowA           580k {\tiny (1.28x)}\hspace{-.5em} & \rowA           568k {\tiny (1.26x)}\hspace{-.5em} & \rowA           568k {\tiny (1.25x)}\hspace{-.5em} & \rowA           493k {\tiny (1.16x)}\hspace{-.5em} & \rowA           419k {\tiny (1.14x)}\hspace{-.5em} & \rowA           420k {\tiny (1.15x)}\hspace{-.5em} & \rowA           530k {\tiny (1.17x)}\hspace{-.5em} \\
 & \rowB OSSL fe-51+\textbf\cryptopt                & \rowB            asm & \rowB           709k {\tiny (1.24x)}\hspace{-.5em} & \rowB           536k {\tiny (1.26x)}\hspace{-.5em} & \rowB           528k {\tiny (1.24x)}\hspace{-.5em} & \rowB           512k {\tiny (1.22x)}\hspace{-.5em} & \rowB           571k {\tiny (1.27x)}\hspace{-.5em} & \rowB           579k {\tiny (1.28x)}\hspace{-.5em} & \rowB           571k {\tiny (1.26x)}\hspace{-.5em} & \rowB           508k {\tiny (1.19x)}\hspace{-.5em} & \rowB           399k {\tiny (1.09x)}\hspace{-.5em} & \rowB           398k {\tiny (1.09x)}\hspace{-.5em} & \rowB           524k {\tiny (1.16x)}\hspace{-.5em} \\
 & \rowA OSSL fe-64 ots~\cite{openssl}              & \rowA            asm & \rowA  \textbf{572k {\tiny (1.00x)}}\hspace{-.5em} & \rowA           444k {\tiny (1.04x)}\hspace{-.5em} & \rowA           444k {\tiny (1.04x)}\hspace{-.5em} & \rowA           431k {\tiny (1.02x)}\hspace{-.5em} & \rowA           462k {\tiny (1.02x)}\hspace{-.5em} & \rowA           460k {\tiny (1.02x)}\hspace{-.5em} & \rowA           464k {\tiny (1.03x)}\hspace{-.5em} & \rowA           465k {\tiny (1.09x)}\hspace{-.5em} & \rowA           411k {\tiny (1.12x)}\hspace{-.5em} & \rowA           412k {\tiny (1.13x)}\hspace{-.5em} & \rowA           455k {\tiny (1.01x)}\hspace{-.5em} \\
 & \rowB OSSL fe-64+\textbf\cryptopt                & \rowB            asm & \rowB           594k {\tiny (1.04x)}\hspace{-.5em} & \rowB           455k {\tiny (1.07x)}\hspace{-.5em} & \rowB           460k {\tiny (1.08x)}\hspace{-.5em} & \rowB           456k {\tiny (1.08x)}\hspace{-.5em} & \rowB           482k {\tiny (1.07x)}\hspace{-.5em} & \rowB           489k {\tiny (1.08x)}\hspace{-.5em} & \rowB           480k {\tiny (1.06x)}\hspace{-.5em} & \rowB           446k {\tiny (1.05x)}\hspace{-.5em} & \rowB           387k {\tiny (1.05x)}\hspace{-.5em} & \rowB           392k {\tiny (1.07x)}\hspace{-.5em} & \rowB           461k {\tiny (1.02x)}\hspace{-.5em} \\
 & \rowA HACL*~fe-64~\cite{hacl}                    & \rowA            asm & \rowA           588k {\tiny (1.03x)}\hspace{-.5em} & \rowA           496k {\tiny (1.16x)}\hspace{-.5em} & \rowA           493k {\tiny (1.16x)}\hspace{-.5em} & \rowA           473k {\tiny (1.12x)}\hspace{-.5em} & \rowA  \textbf{451k {\tiny (1.00x)}}\hspace{-.5em} & \rowA  \textbf{451k {\tiny (1.00x)}}\hspace{-.5em} & \rowA  \textbf{452k {\tiny (1.00x)}}\hspace{-.5em} & \rowA  \textbf{426k {\tiny (1.00x)}}\hspace{-.5em} & \rowA  \textbf{367k {\tiny (1.00x)}}\hspace{-.5em} & \rowA  \textbf{366k {\tiny (1.00x)}}\hspace{-.5em} & \rowA  \textbf{452k {\tiny (1.00x)}}\hspace{-.5em} \\

\midrule
\multirow{4}{*}[+.6em]{\rotatebox[origin=c]{90}{\centering secp256k1}}
 & \rowA libsecp256k1~\cite{libsecp256k1}           & \rowA            asm & \rowA           719k {\tiny (1.07x)}\hspace{-.5em} & \rowA           567k {\tiny (1.05x)}\hspace{-.5em} & \rowA           567k {\tiny (1.05x)}\hspace{-.5em} & \rowA           560k {\tiny (1.06x)}\hspace{-.5em} & \rowA           565k {\tiny (1.03x)}\hspace{-.5em} & \rowA           563k {\tiny (1.02x)}\hspace{-.5em} & \rowA           564k {\tiny (1.02x)}\hspace{-.5em} & \rowA           539k {\tiny (1.07x)}\hspace{-.5em} & \rowA           436k {\tiny (1.04x)}\hspace{-.5em} & \rowA           439k {\tiny (1.06x)}\hspace{-.5em} & \rowA           547k {\tiny (1.04x)}\hspace{-.5em} \\
 & \rowB libsecp256k1~\cite{libsecp256k1}           & \rowB              C & \rowB  \textbf{671k {\tiny (1.00x)}}\hspace{-.5em} & \rowB  \textbf{540k {\tiny (1.00x)}}\hspace{-.5em} & \rowB  \textbf{539k {\tiny (1.00x)}}\hspace{-.5em} & \rowB           529k {\tiny (1.01x)}\hspace{-.5em} & \rowB           562k {\tiny (1.02x)}\hspace{-.5em} & \rowB           563k {\tiny (1.02x)}\hspace{-.5em} & \rowB           562k {\tiny (1.02x)}\hspace{-.5em} & \rowB           505k {\tiny (1.00x)}\hspace{-.5em} & \rowB           437k {\tiny (1.04x)}\hspace{-.5em} & \rowB           435k {\tiny (1.05x)}\hspace{-.5em} & \rowB           530k {\tiny (1.01x)}\hspace{-.5em} \\
 & \rowA libsecp256k1+\textbf\cryptopt (Fiat)       & \rowA            asm & \rowA           705k {\tiny (1.05x)}\hspace{-.5em} & \rowA           548k {\tiny (1.02x)}\hspace{-.5em} & \rowA           550k {\tiny (1.02x)}\hspace{-.5em} & \rowA           533k {\tiny (1.01x)}\hspace{-.5em} & \rowA  \textbf{549k {\tiny (1.00x)}}\hspace{-.5em} & \rowA  \textbf{550k {\tiny (1.00x)}}\hspace{-.5em} & \rowA           554k {\tiny (1.00x)}\hspace{-.5em} & \rowA  \textbf{504k {\tiny (1.00x)}}\hspace{-.5em} & \rowA  \textbf{419k {\tiny (1.00x)}}\hspace{-.5em} & \rowA  \textbf{416k {\tiny (1.00x)}}\hspace{-.5em} & \rowA  \textbf{527k {\tiny (1.00x)}}\hspace{-.5em} \\
 & \rowB libsecp256k1+\textbf\cryptopt (Opt)        & \rowB            asm & \rowB           718k {\tiny (1.07x)}\hspace{-.5em} & \rowB           542k {\tiny (1.00x)}\hspace{-.5em} & \rowB           540k {\tiny (1.00x)}\hspace{-.5em} & \rowB  \textbf{527k {\tiny (1.00x)}}\hspace{-.5em} & \rowB           551k {\tiny (1.00x)}\hspace{-.5em} & \rowB           556k {\tiny (1.01x)}\hspace{-.5em} & \rowB  \textbf{553k {\tiny (1.00x)}}\hspace{-.5em} & \rowB           509k {\tiny (1.01x)}\hspace{-.5em} & \rowB           421k {\tiny (1.01x)}\hspace{-.5em} & \rowB           418k {\tiny (1.01x)}\hspace{-.5em} & \rowB           528k {\tiny (1.00x)}\hspace{-.5em} \\
\bottomrule
\end{tabular}

        \captionof{table}{\label{f:fulltable2}Cost of scalar multiplication (in cycles) of different implementations benchmarking on different machines.}%
    \end{landscape}
    \clearpage%
}
\fi

\end{document}